\documentclass[12pt]{article}
\usepackage{amsfonts}
\usepackage{amsmath}
\usepackage{amssymb}
\usepackage{array}
\usepackage{bigints}
\usepackage{booktabs}
\usepackage[nosort]{cite}
\usepackage{color}
\usepackage{dsfont}
\usepackage{float}
\usepackage{framed}
\usepackage{graphicx}
\usepackage{indentfirst}
\usepackage{mathrsfs}
\usepackage{multirow}
\usepackage{pdflscape}
\usepackage{setspace}
\usepackage{subdepth}
\usepackage{subfig}
\usepackage{titlesec}
\usepackage[dotinlabels]{titletoc}
\usepackage{wrapfig}
\usepackage[all]{xy}
\usepackage{young}
\usepackage[vcentermath]{youngtab}
\usepackage{relsize}
\usepackage{stackengine}
\usepackage{datetime}

\usepackage{hyperref}
\hypersetup{colorlinks=true}
\hypersetup{linkcolor=black}
\hypersetup{citecolor=black}
\hypersetup{urlcolor=black}

\numberwithin{equation}{section}

\usepackage{verbatim}

\newcommand{\be}{\begin{equation}}
\newcommand{\ee}{\end{equation}}
\def\({\left(} \def\){\right)}
\def\[{\left[} \def\]{\right]}

\title{}
\def\sgn{\text{sgn}}

\def\sL{\boldsymbol{L}}
\def\sM{\boldsymbol{M}}

\def\mF{\mathcal{F}}
\def\mJ{\mathcal{J}}
\def\tG{\tilde{G}}
\def\tS{\tilde{\Sigma}}
\def\al{\alpha}
\def\mK{\mathcal{K}}
\def\tr{\text{tr}}
\def\Pf{\text{Pf}}
\def\mO{\mathcal{O}}
\def\mC{\mathcal{C}}

\newcommand{\bea}{\begin{eqnarray}}
\newcommand{\eea}{\end{eqnarray}}

\usepackage[left=2.5cm,right=2.5cm,top=2.5cm,bottom=2.5cm]{geometry}
\titleformat{\section}{\normalfont\bfseries}{\thesection.}{4pt}{}
\titlespacing{\section}{0pt}{22pt}{6pt}

\titleformat{\subsection}{\normalfont\itshape}{\thesubsection.}{4pt}{}
\titlespacing{\subsection}{0pt}{18pt}{6pt}

\titleformat{\subsubsection}{\normalfont\itshape}{\thesubsubsection.}{4pt}{}
\titlespacing{\subsubsection}{0pt}{16pt}{6pt}

\def\ie{\begin{equation}\begin{aligned}}
\def\fe{\end{aligned}\end{equation}}

\def\tilde{\widetilde}

\def\bar{\overline}

\def\1{{\mathds 1}}

\DeclareFontShape{OT1}{cmr}{mx}{n}%
    {<->cmr10}{}
\newcommand{\mytitlefont}{\fontseries{mx}\selectfont}
\DeclareMathAlphabet{\titlemath}{OT1}{cmr}{mx}{n}

\begin{document}


\begin{titlepage}

\begin{center}

~\\[2cm]

{\fontsize{20pt}{0pt} \mytitlefont A Generalization of Sachdev-Ye-Kitaev}

~\\[0.5cm]

{\fontsize{14pt}{0pt} David J.~Gross and Vladimir Rosenhaus}

~\\[0.1cm]

\it{Kavli Institute for Theoretical Physics}\\ \it{University of California, Santa Barbara, CA 93106}

~\\[0.8cm]

\end{center}

\noindent The SYK model: fermions with a $q$-body, Gaussian-random, all-to-all interaction, is the first of a fascinating new class of solvable large $N$ models. We generalize SYK to include $f$ flavors of fermions, each occupying $N_a$ sites and appearing with a $q_a$ order in the interaction. Like SYK, this entire class of models generically has an infrared fixed point. We compute the infrared dimensions of the fermions, and the spectrum of singlet bilinear operators. We show that there is always a dimension-two operator in the spectrum, which implies that, like in SYK,  there is breaking of conformal invariance and maximal chaos in the infrared four-point function of the generalized model. After a disorder average, the generalized model has a global $O(N_1) \times O(N_2) \times \ldots\times O(N_f)$ symmetry: a subgroup of the $O(N)$ symmetry of SYK; thereby giving a richer spectrum. We also elucidate aspects of the large $q$ limit and the OPE, and solve $q=2$ SYK at finite $N$. 
\vfill


\end{titlepage}


\tableofcontents

~\\[2cm]

\section{Introduction}

The Sachdev-Ye  model \cite{SY}, as recently revived and simplified by Kitaev \cite{Kitaev},  possesses, for large $N$,  three remarkable properties:  conformal invariance in the infrared, solvability, and maximal chaos. While there are models that contain some of these properties, SYK is the first to have all three, as was recognized by Kitaev in a series of incredibly insightful seminars \cite{Kitaev}.

Broadly speaking, until recently  two classes of large $N$ theories have been studied: matrix models and vector models, in which the dynamical variables transform in the adjoint or fundamental representation of a local or global $SU(N)$ or $O(N)$ symmetry group, respectively.  Matrix models are closely related to string theories \cite{t1973planar, Polyakov:1981rd, Gross:1993hu, DiFrancesco:1993cyw, Banks:1996vh, Polyakov:1997tj}, with the most concrete realization being the duality between supersymmetric gauge theories and string theory in Anti-de Sitter space \cite{Maldacena:1997re}. $\mathcal{N} =4$ super Yang-Mills is conformally invariant, and at large 't Hooft coupling the bulk gravity has black holes, so it should be maximally chaotic. However, it is not easily solvable. Vector models also have a long history and recently have been shown to be dual to interesting gravity theories. The critical $O(N)$ vector model is conformally invariant and solvable, and the bulk dual is higher spin Vasiliev theory \cite{Klebanov:2002ja, Vasiliev:1999ba}.
However, it is integrable  for large ${N}$,  so it is not likely to be chaotic. 
Roughly speaking, matrix models are too difficult to be explicitly solvable, while vector models are too simple to have the same rich properties. One would like a model that lies in between: one that is sufficiently complicated to be chaotic, while still simple enough to allow for direct analytic calculations for strong coupling. SYK is such a model.

At large $N$, the dominant Feynman diagrams for matrix models are planar diagrams, whereas the dominant diagrams for vector models are bubble diagrams. The SYK model is dominated by a new class of Feynman diagrams, which have been referred to as sunset, or watermelon, diagrams. The SYK model may be just one example out of a much broader  and new class of models. Past studies of large $N$ models have been extremely fruitful for understanding  both quantum field theories and string theories. One may hope that the study  of SYK-like models will also prove productive. 

SYK is a quantum mechanics model, living in  $0+1$ dimensions. While two-dimensional CFTs have been extensively studied and categorized, one dimensional CFTs have not. In fact, it has been argued that $0+1$ dimensional CFTs with nontrivial dynamics do not actually exist \cite{Jensen:2011su}. SYK confirms this: the four-point function breaks the $SL(2,R)$ conformal invariance \cite{Kitaev, PR, MS}, consistent with holographic studies of AdS$_2$ \cite{AP}. It appears that in one dimension a theory can at best  only be ``nearly'' conformally invariant. In SYK the breaking of conformal invariance, to leading order in $1/N$,  is confined to a single dimension-two operator appearing in the OPE, so the power of conformal invariance is still largely applicable.

The SYK model consists of $N \gg 1$ Majorana fermions $\chi_i$, with a $q$-body Hamiltonian with quenched disorder, 
\be \label{SYKI}
H=\sum_{i_1, \ldots, i_q} J_{i_1, \ldots, i_q}\,\, \chi_{i_1} \chi_{i_2} \cdots \chi_{i_q}~.
\ee
The model has qualitatively similar properties for any choice of even $q\geq 4$. The couplings $J_{i_1, \ldots,i_q}$ are independently chosen from a Gaussian, $O(N)$ invariant, distribution with zero mean and a variance proportional to $J^2 N^{1-q}$. When evaluating observables, say correlation functions, a disorder average is performed at the end of the calculation. For the purposes of  correlation functions, at large $N$, the model is self-averaging for $q>2$: randomly chosen, but fixed, $J_{i_1, \ldots, i_q}$ give the same results as disorder averaged $J_{i_1, \ldots, i_q}$. One can alternatively think of the $J_{i_1, \ldots, i_q}$ as nearly static  free bosonic fields; at leading order in $1/N$, this gives the same connected correlation functions \cite{MPRS}, and furthermore, allows one to gauge the $O(N)$ symmetry \cite{Polchinski}. To leading order in $1/N$ the fermions are non-interacting, and  the two-point function of the fermions satisfies a simple integral equation which can be explicitly solved near the infrared fixed point. The fermions  start with dimension $0$ in the UV, and flow to dimension $\Delta = 1/q$ in the IR. 

After the disorder average, the dynamics is invariant under an  $O(N)$ global symmetry, $ \chi_i\rightarrow O_{{ij} }\chi_j$, with $ O O^{T}=1$, much like a vector model. The bilinear, primary, fermion operators,   singlets under $O(N)$, are schematically $\sum_{i=1}^N \chi_i\, \partial_{\tau}^{2n+1} \chi_i$. In the UV, these operators have dimension $2n+1$. In the IR, the dimensions receive an order-one shift for small $n$, and approach $2\Delta+ 2n+1$ asymptotically for large $n$. The standard AdS/CFT dictionary relates the dimensions of CFT single-trace operators for matrix theories, or bilinear singlet operators for vector models, to the masses of particles in the bulk dual. This would imply that the SYK dual has a tower of particles in the bulk, with masses, in units of the AdS radius, roughly spaced by two. This spectrum differs from $\mathcal{N}=4$/$AdS_5 \times S^5$ duality where for large 't Hooft coupling only a small number of massless modes survive, or vector model/Vasiliev duality, where a tower of massless modes appears in the bulk. In \cite{MS} it was noted that the bulk dual of SYK might be a string theory with the string scale comparable to the AdS radius, and thus non-local or stringy. But what the  dual of SYK is, and the extent to which it is nonlocal, remains an open problem.

The goal of this paper is to generalize the SYK model. We would like to understand how large the class of such models is, and which features are generic  and which are special to SYK. This paper will not add anything new to the bulk interpretation of SYK, but the dual bulk theory, whatever it is, should be able to incorporate this more general class of models.

Two seemingly important ingredients in SYK are: (a) $0+1$ dimensions, where the fermions are dimensionless, thereby ensuring that any product of fermions is a relevant perturbation, and (b) quenched disorder, which plays an important role in the solvability at large $N$. The generalization we explore is one in which there are $f$ flavors of fermions, $\chi_{i}^a$, where $i=1 \ldots N_{a}$ and $ a=1\ldots f$, with a Hamiltonian, 
\be \label{GRI}
H= \sum_I J_{I} (\chi_{i_1}^1 \cdots \chi_{i_{q_1}}^1)\cdots (\chi_{j_1}^f \cdots \chi_{j_{q_f}}^f)~,
\ee
where $I$ is a collective site index, and the subscript on the fermion is the site while the superscript is the flavor. The number of sites, $N_a$, for each fermion, as well as the order of the interaction, $q_{a}$,  can depend on the flavor $a$, as long as  ${N_{a}/N}$ remains finite as $N = \sum_{a} N_{a} \to \infty$. 

In Sec.~\ref{Sec2pt} we derive the Schwinger-Dyson equation for the two-point functions of the fermions, and find that the model (\ref{GRI}) generically has an IR fixed point. While the IR  dimension for SYK  was $\Delta = 1/q$, for the generalized model (\ref{GRI})  a set of $f$  transcendental equations determine the  dimensions $\Delta_a$. In the limit of large $q_a$, these have simple analytic solutions. Furthermore, for large $q_a$ one only needs to sum a particular subset of Feynman diagrams, thus yielding an explicit expression for the two-point function and the spectral function. 

In Sec.~\ref{Sec4pt} we study the spectrum of composite operators. After the disorder average, the generalized model (\ref{GRI}) has an $O(N_1)\times O(N_2) \times \cdots \times O(N_f)$ symmetry. The singlet bilinear operators are schematically  $\sum_{i=1}^{N_{a}} \chi_i^a\, \partial_{\tau}^{1+ 2n} \chi_i^a$ for any $a \in \{1, \ldots, f\}$. So we expect there to be $f$ towers of operators. We derive equations determining the IR dimensions of these operators. We prove that for any choice of parameters: $f$, $N_a$'s, $q_a$'s, there is always a dimension-two operator in the spectrum. In SYK, the dimension-two operator is responsible for both the breaking of conformal symmetry in the four-point function and for maximal chaos. The same properties hold for the generalized model. 

An instructive case to study is the generalized model with all $N_a$ equal to $N/f$ and all $q_a$ equal to $q$. It has the symmetry $O(N/f) \times \cdots \times O(N/f)$. The spectrum contains a tower identical to that of SYK with a $q f$ body interaction, along with a new tower that appears with a  degeneracy of $f-1$. Indeed, this model is similar to SYK with $N$ fermions  and a $q f$ body interaction, but the full $O(N)$ symmetry is broken and consequently more singlet operators exist, allowing for a richer model.
 
In Appendix~\ref{free} we consider the path integral for the generalized model. This provides an alternate way of computing the correlation functions, with the saddle point giving the Schwinger-Dyson equations for the two-point functions, and the leading $1/N$ fluctuations about the saddle giving the four-point function. In Appendix~\ref{AppendixScalar} we consider (\ref{GRI}) with an additional scalar; a special case of this includes supersymmetric SYK \cite{Anninos:2016szt}.
 Finally, Appendix~\ref{AppendixA} solves SYK for $q=2$ at finite $N$. SYK for $q=2$ is like $N$ fermions with a random mass matrix; the randomness makes it nontrivial, though it is less interesting than $q\geq4$. This appendix can be read independently of the rest of the paper.

\section{Two-Point Function} \label{Sec2pt}

\subsection{SYK}
Let us recall the SYK model \cite{Kitaev}.~\footnote{For recent studies of SYK, see \cite{PR, MS, Sachdev15, Anninos:2016szt, You:2016ldz, Jensen:2016pah, Fu:2016yrv, Jevicki:2016bwu, Jevicki:2016ito, Bag, Dan}. For some related studies of AdS$_2$ and conformal symmetry breaking see \cite{MSY, Engelsoy:2016xyb, Almheiri:2016fws, Cvetic:2016eiv, Radicevic:2016kpf}. For earlier studies of a holographic interpretation of the SY model, see \cite{Sachdev:2010um, Sachdev:2010uj}. While this paper was being completed, \cite{Gu:2016oyy} appeared, which considers a higher dimensional generalization of SYK.} It contains $N$ Majorana fermions with the anticommutation relation $\{\chi_i, \chi_j\} = \delta_{i j}$. The action is, 
\be \label{SYK}
S= \int d \tau\,  \( \frac{1}{2} \sum_{i=1}^N\chi_i \frac{d }{d \tau} \chi_i\, +\, \frac{(i)^{\frac{q}{2}}}{q!} \sum_{i_1, \ldots, i_q=1}^N J_{ i_1 i_2 \ldots i_q} \chi_{i_1} \chi_{i_2}\cdots \chi_{i_q} \)~,
\ee
where the coupling $J_{i_1, \ldots, i_q}$ is totally antisymmetric and, for each $i_1, \ldots, i_q$, is chosen from a Gaussian ensemble. The two-point function of the $J_{i_1, \ldots, i_q}$ is taken to be,
\be \label{disA}
\frac{1}{(q-1)!} \sum_{i_2, \ldots, i_q=1}^{N}\langle J_{i_1 i_2 \ldots i_q}  J_{i_1 i_2 \ldots i_q}\rangle= J^2~.
\ee
At leading order in $1/N$, (\ref{disA}) is equivalent to the simpler normalization,
\be \label{SYKdis}
\langle J_{i_1 i_2 \ldots i_q}  J_{i_1 i_2 \ldots i_q}\rangle= (q-1)! \frac{J^2}{N^{q-1}}~.
\ee
The particular scaling with $N$ in the choice (\ref{SYKdis}) is in order to obtain a nontrivial large $N$ limit, while the other factors are for convenience. One can consider SYK for any even $q\geq 2$, with  $q=4$ being the prototypical case \cite{Kitaev}. At $q=\infty$ there are some simplifications \cite{MS}. The case $q=2$ is simplest, and is equivalent to an $O(N)$ vector fermion with a random mass matrix, although in many ways it is qualitatively different from the $q>2$ models. We solve the $q=2$ SYK at finite $N$ in Appendix \ref{AppendixA}.

\begin{figure}[t]
\centering 
\includegraphics[width=1.3in]{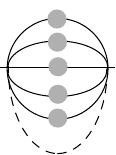}
\caption{The self-energy (\ref{SD2}) for a fermion in SYK (\ref{SYK}). The figure is for $q=6$. The solid line with a filled circle is the two-point function. The dashed line is the disorder.} \label{Fig2pt}
\end{figure} 

At zero coupling, the Euclidean two-point function $\langle T \chi_i(\tau) \chi_j(0)\rangle \equiv G(\tau) \delta_{i j}$ is given by,
\be \label{Gweak}
G_0(\tau) = \frac{1}{2} \sgn(\tau)\ , \ \ \ \ G_0(\omega) = \frac{i}{\omega}~,
\ee
where the factor $\sgn(\tau)$ ($\sgn(\tau) = 1$ for $\tau>0$ and $\sgn(\tau) = -1$ for $\tau<0$) accounts for the fermion anticommutation. 
To leading order in $1/N$, the  Schwinger-Dyson equations for the two-point function drastically simplify and are given by,
\bea \label{SD}
G(\omega)^{-1} &=&G_0(\omega)^{-1}- \Sigma(\omega)= -i \omega - \Sigma(\omega)~, \\ \label{SD2}
\Sigma(\tau) &=& J^2 G(\tau)^{q-1}~.
\eea
The first of these, (\ref{SD}), is the standard equation expressing the two-point function in terms of the one-particle irreducible self-energy $\Sigma(\omega)$. The second equation, which is written in position space, is a special feature of SYK (see Fig.~\ref{Fig2pt}). At leading order in $1/N$, the only diagrams that survive are nested sunset diagrams; all others are suppressed by some power of $1/N$. These equations  can be combined into a single integral equation; however, an analytic solution to this equation is not known. 
At strong coupling, $|J \tau| \gg 1$ (equivalently, the infrared limit), one can drop the $i\omega$ in (\ref{SD}), to get, 
\be \label{SDIR}
G(\omega) \Sigma(\omega) = -1~,\ \ \ \ \ \ \
\Sigma(\tau) = J^2 G(\tau)^{q-1}~.
\ee
One can verify that,
\be \label{Gstrong}
G(\tau) = b \frac{\sgn(\tau)}{|J \tau|^{2\Delta}}
\ee
is a solution to (\ref{SDIR}) provided one takes, 
\be \label{DeltaSYK}
\Delta= \frac{1}{q}~, \ \ \ \ \ \ \ \ b^{q} =\frac{1}{2\pi}\(1 - 2 \Delta\) \tan \pi \Delta~.
\ee
The Fourier transform of $G(\tau)$, given in (\ref{Gstrong}), is useful in verifying this, 
\be  \label{FT}
G(\omega) = \int d \tau\, e^{i \omega \tau} G(\tau) =b\,  \psi(\Delta) J^{- 2\Delta} |\omega|^{2\Delta -1} \sgn(\omega),
\ee
where we defined,
\be \label{psiDelta}
\psi (\Delta) \equiv 2 i \cos (\pi \Delta) \Gamma(1-2\Delta) =  2 i \sqrt{\pi}\, 2^{-2\Delta} \frac{\Gamma(1- \Delta)}{\Gamma(\frac{1}{2} + \Delta)}~.
\ee
What is special to SYK is that the  IR Schwinger-Dyson equations (\ref{SDIR}) are invariant under reparameterization of time, $\tau\rightarrow f(\tau)$, with the propagator transforming as  $G(\tau_1 - \tau_2) \rightarrow f'(\tau_1)^{\Delta} f'(\tau_2)^{\Delta}\, G(f(\tau_1) - f(\tau_2))$. Therefore, although (\ref{Gstrong}) is at zero temperature, we can easily construct the finite-temperature two-point function by mapping the real line to a circle \cite{Kitaev, PGKS, PG, GPS}.

\subsection{A Generalization of SYK} \label{Sec:Gen}
The model we introduce is a generalization of SYK (\ref{SYK}). It contains $f$ flavors of fermions, with $N_a$ fermions of flavor $a$, each appearing $q_a$ times in the interaction, so that the Hamiltonian couples $\mathrm{q}=\sum_{a=1}^f q_a$ fermions together. We continue to let the subscript on the fermion $\chi_i^a$ denote the site $i\in \{1,\ldots, N_a\}$, while the superscript $a$ will now denote the flavor $a\in\{1,\ldots, f\}$. Explicitly, the action is, 
\be \label{GFM}
S= \int d \tau\,  \( \frac{1}{2}\sum_{a =1}^f \sum_{i=1}^{N_a}\chi_i^a \frac{d }{d \tau} \chi_i^a +\frac{(i)^{\frac{\mathrm{q}}{2}}}{\prod_{a=1}^f q_a!} \sum_I J_{I}( \chi_{i_1}^1 \cdots \chi_{i_{q_1}}^1)\cdots (\chi_{j_1}^f \cdots \chi_{j_{q_f}}^f) \)~,
\ee
where $I$ is a collective index, $I = i_1,\ldots, i_{q_1}, \ldots, j_{1},\ldots, j_{q_f}$. The coupling $J_{I}$ is antisymmetric under permutation of indices within any one of the $f$ families, and  is drawn from a Gaussian distribution, 
\be \label{PJI}
P[J_I] \propto \exp\( -\frac{ \sum_I J_I^2}{2\langle J_I J_I\rangle}\)~,
\ee
where the disorder average $\langle J_I J_I\rangle$ is given by 
\be  \label{DisAvgGF}
\langle J_I J_I\rangle =J^2  \frac{\sum_{a=1}^f N_a}{\prod_a N_a^{q_a}} \prod_a (q_a - 1)!~.
\ee
It will be convenient to make the following definitions, 
\be \label{GFMd}
N \equiv \sum_{a=1}^f N_a~,  \ \ \ \ \ \kappa_k = \frac{N_k}{N}~, \ \ \ \ \ Q_k \equiv \prod_{a\neq k} q_a~.
\ee
The class of models (\ref{GFM}) for large $N$ is characterized by  $f-1$ independent continuous parameters $0<\kappa_k<1$, as well as the $q_k$, which can be any positive integers provided that their sum is even.  After the disorder average, SYK (\ref{SYK}) has $O(N)$ symmetry, while in the generalized model (\ref{GFM}) the symmetry is broken to the subgroup $O(N_1) \times O(N_2) \times \cdots \times O(N_f)$. 

In the SYK model (\ref{SYK}), one can generalize the action to have multiple interaction terms, with different $q$, each coming with its own independent disorder $J_{i_1\ldots i_q}$. This sum of SYK Hamiltonians is just as solvable as SYK, however it is not especially interesting since in the IR the term with smallest $q$ will be dominant. In the model (\ref{GFM}), one can also consider generalizing the action to include sums of interaction terms. However, now the IR can be more interesting, since there can be multiple terms (with the same total  $\mathrm{q}$) that are equally important in the IR. 

\subsubsection*{Two-Point Function}
\begin{figure}[t]
\centering 
\includegraphics[width=1.3in]{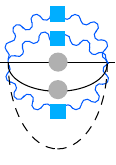}
\caption{The self-energy (\ref{SigGen}) for a fermion of flavor $k$ in the generalized model (\ref{GFM}). The figure is for two flavors with $q_1 = q_2 = 3$. This is, for instance, the self-energy for fermion of flavor $1$. The black line with a filled circle is the two-point function for the fermion of flavor $1$, while the blue wavy line with a filled square is the two-point function for the flavor $2$ fermion.} \label{2p2t}
\end{figure}
The free two-point function for each $\chi_i^a$, is again given by (\ref{Gweak}). Away from the UV it will continue to be the case that the two-point function is diagonal in flavor and site space. 
Denoting the two-point function for the flavor $k$ fermion by $G_k(\tau)$, the self-energy for the flavor $k$ fermion is (see Fig.~\ref{2p2t}), \footnote{The factors are as follows. The factor $\(\prod q_a!\)^2$ comes from the square of the prefactor of the interaction term in (\ref{GFM}). There is a factor of $q_k$ from the number of contractions with the ingoing fermion, and another $q_k$ with the outgoing fermion, and a $(q_k-1)!$ from the contraction of the remaining flavor $k$ fermions amongst themselves. There is also a factor of $q_a!$ from contractions of flavor $a$ fermions, for all the other flavors.}
\be \label{SigGen}
\Sigma_k(\tau) =\langle J_I J_I\rangle \frac{q_k^2 (q_k-1)! \prod_{a\neq k} q_a! }{\(\prod q_a!\)^2} \(N_k G_k(\tau)\)^{q_k - 1} \prod_{ a\neq k} \(N_a G_a(\tau)\)^{q_a}\, ~.
\ee
Making use of (\ref{DisAvgGF}, \ref{GFMd}), this simplifies to,
\be \label{SigGen2}
\Sigma_k(\tau) =J^2 \frac{1}{\kappa_k Q_k} \frac{1}{G_{k}(\tau)} \prod_{a} G_a(\tau)^{q_a}~.
\ee
An alternative way to obtain (\ref{SigGen2}) is by performing the replica trick to do the disorder average, introducing mean fields, integrating out the fermions, and taking the large $N$ saddle point; see Appendix~\ref{free}. For one flavor, (\ref{SigGen2}) reduces to the SYK expression for the self-energy (\ref{SD2}). 

We first determine whether there is an IR fixed point and, if so,  the  IR dimension $\Delta_k$  of the fermions of flavor $k$. In the IR, the two-point function should take the form,
\be \label{Gkto}
G_k(\tau) = b_k \frac{\sgn(\tau)}{|J \tau|^{2 \Delta_k}}~, \ \ \ \  \ \ \ \ G_k(\omega) = b_k \psi(\Delta_k)\, J^{- 2\Delta_k}\, |\omega|^{2\Delta_k -1} \sgn(\omega) ~.
\ee
To find the normalization $b_k$ and dimension $\Delta_k$, we first insert the above ansatz into (\ref{SigGen2}) and take the Fourier transform,~\footnote{One should not confuse the usage of $\Sigma$ as the self-energy with the usage of $\Sigma$ as a sum.}
\be \label{Skto}
\Sigma_k(\omega) = J  \sgn(\omega) \Big|\frac{\omega}{J}\Big|^{\sum\limits_{a=1}^f 2 \Delta_a q_a -2 \Delta_k -1}  \frac{\prod_{a} b_a^{q_a}}{b_k \kappa_k Q_k} \, \psi\(\sum_{a=1}^f \Delta_a q_a - \Delta_k\)~.
\ee
Inserting (\ref{Skto}) and (\ref{Gkto}) into the IR limit of (\ref{SD}), $\Sigma_k(\omega) G_k(\omega) =-1$, gives, 
\bea\label{DimSumG}
1&=& \sum_{a=1}^f \Delta_a q_a~, \\  \label{GenNorm1}
 \prod_{a=1}^f b_a^{q_a} &=& \frac{ - \kappa_k Q_k}{\psi(\Delta_k) \psi(1- \Delta_k)}~.
\eea
The first equation is just the statement that the IR dimension of the coupling $J_I$ is zero. 
Simplifying the second gives,
\be \label{GenNorm2}
\prod_{a=1}^f b_a^{q_a} = \frac{  \kappa_k Q_k}{2\pi} (1-2\Delta_k ) \tan \pi \Delta_k~.
\ee
Equating all the (\ref{GenNorm2}), for $k$ ranging from $1$ to $f$, gives $f-1$ equations. Combined with (\ref{DimSumG}), for any given choices of $\kappa_k$ and $q_k$, we have a set of $f$  equations for the $f$ unknown dimensions $\Delta_k$.~\footnote{These equations generically have solutions. The case of $q_1=1$ appears to be exceptional.  For instance, taking two flavors with  $q_1=1$, $q_2=3$, there is no solution for $\kappa_1<\frac{1}{10}$.}  These equations have simple solutions in the limit of $q_a \gg 1$, as we show in the next section.

\subsection{Large $q_k$} \label{LARGEQ}
If  the number of fermions of flavor $k$ appearing in the interaction (\ref{GFM}) is large, $q_k \gg 1$, then from (\ref{DimSumG}) we know that $\Delta_k \ll 1$. Let us assume $q_k \gg 1$ for all $k$.  
In this limit, (\ref{GenNorm2}) simplifies to $\prod b_a^{q_a} = \frac{1}{2} \kappa_k Q_k \Delta_k$, with the solution, 
\be \label{LargeQDelta}
\Delta_k = \frac{q_k}{\kappa_k} \frac{1}{\sum_{a=1}^f \frac{q_a^2}{\kappa_a}}~.
\ee
Eq.~\ref{LargeQDelta} shows that a hierarchy in the $q_k$'s for different flavors, or in the $\kappa_k$'s, will lead to a hierarchy in the $\Delta_k$'s. 

The smallness of the dimensions $\Delta_a$ suggests one should be able to solve for the two-point function  at all energies. This was done for SYK at large $q$ in \cite{MS}. Here we perform  an analogous computation for the generalized model in the large $q_k$ limit. The two-point functions are taken to be,
\be \label{LQtwo}
G_k(\tau) = \frac{\sgn(\tau)}{2} e^{\frac{g_k(\tau)}{q_k}} \approx \frac{\sgn(\tau)}{2} \( 1 + \frac{g_k(\tau)}{q_k}+\ldots\)~.
\ee
Taking the Fourier transform, for which we use the shorthand $\mF$, 
\be
G_k(\omega) = \frac{i}{\omega} + \frac{1}{2q_k} \mF(g_k\, \sgn(\tau))+ \ldots.
\ee
Inverting to get $G_k(\omega)^{-1}$, (\ref{SD}) allows us to identify, 
\be
\Sigma_k(\omega) = -\frac{\omega^2}{2 q_k} \mF(g_k\, \sgn(\tau))~, \ \ \ \ \ \Sigma_k(\tau) = \frac{1}{2 q_k} \partial_{\tau}^2 \[g_k(\tau)\, \sgn(\tau)\]~,
\ee
where in the second equation we have done an inverse Fourier transform of the first. Combining with (\ref{SigGen2}) gives, 
\be \label{GFMQ}
\partial_{\tau}^2 \[g_k(\tau) \sgn(\tau)\]= \frac{2 J^2 q_k}{\kappa_k Q_k} \frac{1}{2^{\sum q_a -1}}\, e^{\sum g_a(\tau)}\, \sgn(\tau)~.
\ee
We have such an equation for every $k$. Thus, we can express $g_a$ in terms of $g_k$ for any $a, k$, 
\be \label{gagk}
g_a (\tau) = \(\frac{q_a}{q_k}\)^2 \frac{\kappa_k}{\kappa_a}\, g_k (\tau)~.
\ee
Summing (\ref{GFMQ}) for all $k$ and using (\ref{gagk}) gives, 
\be \label{QSUM}
\partial_{\tau}^2 \sum g_a \sgn(\tau) = 2 \mJ^2\, e^{\sum g_a}\, \sgn(\tau)~, \ \ \ \ \ \ \ \mJ^2 \equiv \frac{2 J^2}{2^{\sum q_a }} \frac{1}{\prod q_a} \sum \frac{q_a^2}{\kappa_a}~.
\ee
where the rescaled $\mJ$  is kept finite in the large $q_k$ limit. 
The solution to (\ref{QSUM}) is easily derived for finite temperature, namely with $g_a(\tau) = g_a(\tau+\beta)$ \cite{MS},
\be \label{egv}
e^{\sum_{a} g_a(\tau)} =\[\frac{\cos \(\frac{\pi v}{2}\)}{\cos \(\frac{\pi v}{2} - \frac{\pi v |\tau|}{\beta}\)}\]^2~, \ \ \ \ \ \ \beta \mJ = \frac{\pi v}{\cos \frac{\pi v}{2}}~,
\ee 
where $v$ is defined implicitly in terms of $\mJ$. 
At zero temperature (\ref{egv}) becomes, 
\be
e^{\sum_{a} g_a(\tau)} = \frac{1}{(1 + \mJ |\tau|)^2}~,
\ee
which combined with (\ref{gagk}) gives, 
\be \label{egq}
e^{\frac{g_k(\tau)}{q_k}} = \frac{1}{(1 + \mJ |\tau|)^{2 \Delta_k}}~,
\ee
where $\Delta_k$ is given by (\ref{LargeQDelta}). Having the exact solution, we can take the IR limit $J |\tau| \gg 1$ to find the individual normalizations of the two-point function (\ref{Gkto}), 
\be
b_k  = \frac{1}{2} \frac{J^{2 \Delta_k}}{\mJ^{2 \Delta_k}}~.
\ee
Recall that solving the IR limit of the Schwinger-Dyson equations only established the product of the normalizations, (\ref{GenNorm2}).

\subsubsection{Graphical Solution}
\begin{figure}[t]
\centering
\subfloat[]{
\includegraphics[width=1.2in]{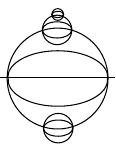} \ \ \ \  \ \ \  \ \  \ \ 
}
\subfloat[]{
\includegraphics[width=1.2in]{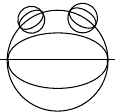} \ \ \ \ \ \ \ \ \ \ \ 
}
\subfloat[]{
\includegraphics[width=1.8in]{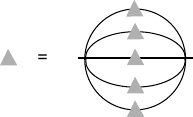}
}
\caption{The two-point function for SYK at large $q$ consists only of  diagrams that split into two trees under a vertical cut, such as the one shown in (a). Diagrams like (b) are suppressed by factors of $1/q$. In (c) we show the recursion relation for the self-energy, represented by a filled triangle. Note that the solid lines are the (free) fermion propagators, and we have suppressed the disorder lines. } \label{LargeQ}
\end{figure}
The two-point function in the large $q$ limit can alternatively be found by summing an appropriate set of Feynman diagrams. We will show how this works in SYK. Due to large $q$ combinatorics, the diagrams contributing to the self-energy that appear most often are like the ones shown in Fig.~\ref{LargeQ} (a), rather than those in Fig.~\ref{LargeQ} (b). The Feynman diagrams that are summed at large $q$ can be characterized as those diagrams that, under a single vertical cut, break up into two tree diagrams. The self-energy can therefore be found recursively, as shown in Fig.~\ref{LargeQ}(c). The equation corresponding to Fig.~\ref{LargeQ}(c) is, 
\be \label{maxCut}
\Sigma(\tau) = J^2 \Big(\int \frac{d\omega}{2\pi}\, e^{- i \omega \tau}\, G_0(\omega)^2 \Sigma(\omega)\Big)^{q-1}~,
\ee
where $G_0(\omega)$ is the free two-point function (\ref{Gweak}). Rearranging (\ref{maxCut}) gives 
\be 
\int d\tau\( J^{-2} \Sigma(\tau)\)^{\frac{1}{q-1}}\, (-\omega^2)\, e ^{i\omega \tau} = \Sigma(\omega)~, \ \ \ \ \ \partial_{\tau}^2 \(J^{-2}\Sigma(\tau)\)^{\frac{1}{q-1}} = \Sigma(\tau)~,
\ee
where the second equation is the inverse Fourier transform of the first. Letting
\be
\Sigma(\tau) = J^2 2^{1-q} \sgn(\tau)\, e^{g(\tau)}~,
\ee
we get,
\be
\partial_{\tau}^2\[ g(\tau) \sgn(\tau)\]= \frac{q-1}{2^{q-2}} J^2\, e^{g(\tau)} \sgn(\tau)~,
\ee
which is (\ref{GFMQ}) for one flavor.

\subsubsection{Spectral Function}
The frequency space two-point function follows from (\ref{LQtwo}, \ref{egq}), 
\be \label{Gqo}
G_k(\omega) = \frac{1}{2} \int_{-\infty}^{\infty} d\tau\, e^{i\omega \tau}\, \frac{\sgn (\tau)}{(1+\mJ |\tau|)^{2\Delta_k}}~.
\ee
Introducing a Schwinger parameter, 
\be
\frac{1}{(1+ \mJ |\tau|)^{2\Delta_k}} = \frac{1}{\Gamma(2\Delta_k)} \int_0^{\infty} d\lambda\, e^{- \lambda(1+ \mJ |\tau|)}\, \lambda^{2\Delta_k - 1}~,
\ee
and performing the $\tau$ integral in Eq.~\ref{Gqo}  gives, 
\be
G_k(\omega) = - \frac{1}{2\Gamma(2 \Delta_k)} \int_{-\infty}^{\infty} d\lambda\,e^{-|\lambda|}\, \frac{|\lambda|^{2\Delta_k - 1}}{i\omega - \lambda \mJ}~.
\ee
The spectral function (as defined in Appendix~\ref{AppendixA} by Eq.~\ref{Gspec}) for the flavor $k$ fermion  is therefore, 
\be
\rho_k(\lambda) = \frac{1}{2 \mJ\, \Gamma(2 \Delta_k)}\, \(\frac{|\lambda|}{\mJ}\)^{2\Delta_k -1}\, e^{- \frac{|\lambda|}{\mJ}}~,
\ee
where $\Delta_k$ is given by (\ref{LargeQDelta}). Since $\Delta_k \ll 1$, this is sharply peaked around small $\lambda$. If there is only one flavor, then $\Delta =1/q$. This spectral function is for $q \gg 1$. For $q=2$, the SYK spectral function is instead a Wigner semicircle (\ref{Wigner}). 

\subsection{Effective Action} \label{EA}
We have so far discussed the model (\ref{GFM}) directly in terms of the fermions, finding the two-point function at large $N$ through study of Feynman diagrams. It is useful to also consider the path integral approach. Employing the replica trick, one can carry out the disorder average, and then integrate out the fermions after the introduction of new (bilocal) fields $\tG_a(\tau_1, \tau_2)$ and $\tS_a(\tau_1, \tau_2)$. The result is (see Appendix~\ref{free}), 
\bea \label{Zeff}
Z &=& e^{-\beta F} = \int D \tS_a\, D \tG_a\, \exp\( - N S_{eff}\), \\ \nonumber
\!\!\!S_{eff}\! \!\!&=&\!\! \!\!-\!\sum_{a=1}^f \kappa_a \log \Pf \(\! \partial_{\tau} - \tS_a\! \) 
\!+\frac{1}{2}\! \int\!\! d\tau_1 d\tau_2\! \[\sum_{a=1}^f \kappa_a\, \tS_a (\tau_1, \tau_2) \tG_a(\tau_1, \tau_2)\! - \frac{J^2}{\prod_a q_a} \prod_{a=1}^f \tG_a(\tau_1, \tau_2)^{q_a}\!\]
\eea
For one flavor, this reduces to the effective action for SYK \cite{Kitaev} (see \cite{SY} for an analogous expression for the SY model, and \cite{Sachdev15} for the Dirac fermion version of SYK). The large $N$ saddle point of the action gives the Schwinger-Dyson equations for the two-point function  found previously from Feynman diagrams. In particular, varying $S_{eff}$ with respect to $\tG_k(\tau_1, \tau_2)$, and assuming time-invariance, gives  (\ref{SigGen2}), while varying with respect to $\tS_k(\tau_1, \tau_2)$ yields (\ref{SD}) for each flavor. The saddle point solutions are denoted by $G_k(\tau_1, \tau_2)$ and $\Sigma_k(\tau_1, \tau_2)$. 

To leading order in $1/N$, the free energy is given by the saddle of (\ref{Zeff}), 
\be \nonumber
\!\!\!\!\!\!- \beta F/N\! =\!   \sum_{a=1}^f \kappa_a \log \Pf \(\partial_{\tau} - \Sigma_a\) 
- \frac{1}{2}\! \int_0^{\beta}\!\! d\tau_1 d\tau_2\! \[\!\sum_{a=1}^f \!\kappa_a\, \Sigma_a (\tau_1, \tau_2) G_a(\tau_1, \tau_2) \! - \frac{J^2}{\prod_a q_a}\! \prod_a\! G_a(\tau_1, \tau_2)^{q_a}\!\]
\ee
Following \cite{MS}, one can differentiate with respect to $J$ to get, 
\be
J \partial_J (- \beta F/N) = \frac{J^2 \beta}{\prod_a q_a}\int_0^{\beta}d\tau\, \prod_a G_a(\tau)^{q_a}~.
\ee
For large $q_a$, $G_a(\tau)$ was found explicitly in Sec.~\ref{LARGEQ}. Also, since the partition function only depends on $\beta J$, it follows that $J\partial_J = \beta \partial_{\beta}$. Thus for large $q_a$, 
\be
\beta\partial_{\beta} (-\beta F/N) = \frac{J^2 \beta}{2^{1 + \sum_a q_a} \prod_a q_a}\int_0^{\beta} d\tau\[\frac{\cos \(\frac{\pi v}{2}\)}{\cos \(\frac{\pi v}{2} - \frac{\pi v |\tau|}{\beta}\)}\]^2~,
\ee 
where $v$ is defined in terms of $\mJ$ in (\ref{egv}). Up to the choice of normalization of the variance of the disorder, $\langle J_I J_I\rangle$, this is the same as for SYK with $N$ fermions and a $\sum_{a=1}^f q_a$ body interaction. So the entropies are also the same. In order to see a distinction, one must study the $1/N$ corrections.

\section{Four-Point Function} \label{Sec4pt}
The SYK model has an $O(N)$ symmetry after the disorder average. The bilinear primary operators that are $O(N)$ invariant are schematically $\sum_{i} \chi_i\, \partial_{\tau}^{1+ 2 n} \chi_i$ for nonnegative integer $n$. In the UV, these have dimension $2 n+1$. The IR dimensions of the operators are computed by summing a class of ladder diagrams. The four-point function of the fermions is then given by a  sum over conformal blocks, one for each of these composite operators. 

For the generalized model (\ref{GFM}), there is an $O(N_1)\times O(N_2) \times \cdots \times O(N_f)$ symmetry after the disorder average, and the invariant operators are schematically  $\sum_i \chi_i^a\, \partial_{\tau}^{1+ 2n} \chi_i^a$ for any $a \in \{1, \ldots, f\}$. So there are now $f$ towers of operators. In this section we compute the IR dimensions of these operators.

\subsection{Dimensions of Composite Operators}  \label{DimComp}
\begin{figure}[t]
\centering
\subfloat[]{
\includegraphics[width=4.5in]{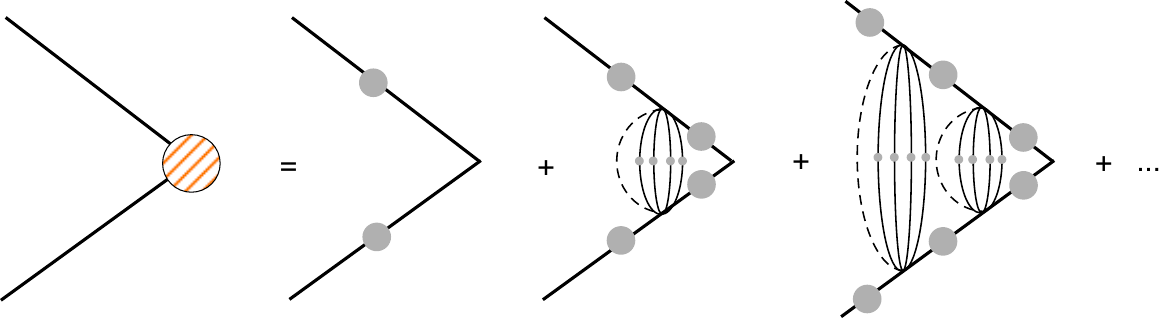} }\ \ \ \ \ \ \ \ \ 
\subfloat[]{
\includegraphics[width=3.3 in]{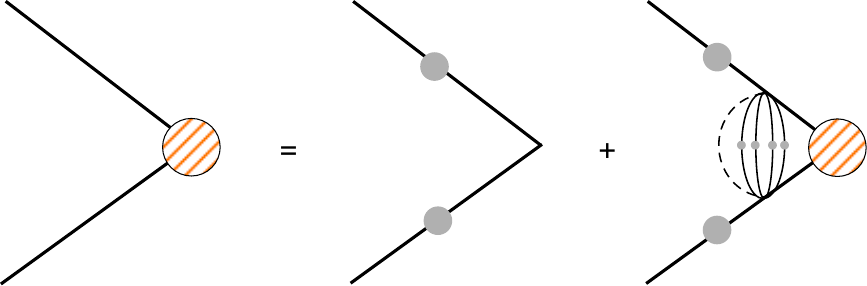}
}\\
\subfloat[]{
\includegraphics[width=1.2 in]{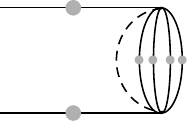} \ \ \ \ \  \ \ \ \  \ \ \ \ \ \ \ \ \ \ \ \ \ \ 
}
\subfloat[]{
\includegraphics[width=1.7 in]{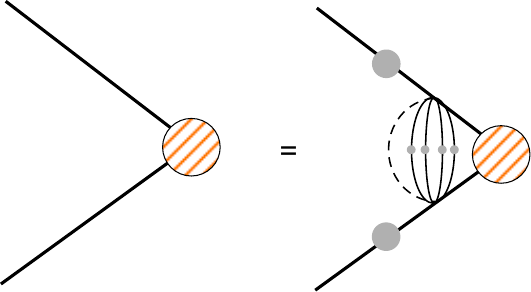}
}
\caption{(a) The diagrams being summed to compute the three-point function $\langle\chi_i(\tau_1) \chi_j(\tau_2) \mO(\tau_0)\rangle$ for $q=6$ SYK. This can be done iteratively, as shown in (b) (see Eq.~\ref{EIG1}), with the kernel shown in  (c) adding rungs to the ladder. In the IR we can simplify (b) to get (d).} \label{BS}
\end{figure}

We begin by reviewing and adding some detail to the computation in \cite{Kitaev} for the IR dimensions of the SYK composite operators. The primary $O(N)$ invariant  bilinear operators are, 
\be \label{O2n1}
\mO_{n} = \sum_{i=1}^N \sum_{k=0}^{2n+1} d_{n k}\, \partial_{\tau}^{k} \chi_i\, \partial_{\tau}^{2n+1 - k} \chi_i~,
\ee
where the coefficients $d_{n k}$ are chosen so that the operators are primary. For instance, 
\be
\mO_1  = \frac{1}{2} \sum_{i=1}^N \partial_{\tau}^2 \chi_i \partial_{\tau} \chi_i - \partial_{\tau}\chi_i \partial_{\tau}^2 \chi_i~.
\ee
The general form of $d_{n k}$ will not be important for us. 

We would like to compute the overlap between the state created by the composite operator $\mO_n$ acting at time $\tau_0$,  and two fermions at times $\tau_1$ and $\tau_2$, respectively. In other words, the  three-point function, 
$\langle \chi_i(\tau_1) \chi_i(\tau_2) \mO(\tau_0)\rangle$, which we will denote by $v(\tau_0; \tau_1, \tau_2)$. If the two fermions just propagated without interacting with each other, this would be found by Wick contractions,
\be \label{GccO}
G^0_{\chi \chi \mO}  = \sum_{i=1}^N\, \sum_{k=0}^{2n+1} d_{ n k }\, \( \partial_{\tau_0}^k G(\tau_2, \tau_0) \partial_{\tau_0}^{2n+1 - k} G(\tau_1, \tau_0) - \partial_{\tau_0}^k G(\tau_1, \tau_0) \partial_{\tau_0}^{2n+1-k} G(\tau_2, \tau_0)\)~.
\ee

Eq.~\ref{GccO} is the first diagram that appears in Fig.~\ref{BS}a. We must also include a sum over all the ladder diagrams in Fig.~\ref{BS}a. 
One can perform the sum by solving the equation (see Fig.~\ref{BS}b), 
\be \label{EIG1}
v(\tau_0; \tau_1, \tau_2) =G^0_{\chi \chi \mO}(\tau_1, \tau_2, \tau_0) +  \int d\tau_3 d\tau_4\, K(\tau_1, \tau_2, \tau_3, \tau_4)\, v(\tau_0; \tau_3, \tau_4)~,
\ee
where the kernel is the operator that adds a single rung (see Fig.~\ref{BS}c),
\be \label{K1flav}
K(\tau_1, \tau_2, \tau_3, \tau_4) = - J^2 (q-1) G(\tau_{13}) G(\tau_{24}) G(\tau_{34})^{q-2}~,
\ee
where $\tau_{ab} \equiv \tau_a - \tau_b$. Letting the composite have  dimension $h$, in the IR the solution to (\ref{EIG1}) will take the form of conformal three-point function, 
\be \label{v012}
v(\tau_0; \tau_1, \tau_2) =\frac{1}{|\tau_1 - \tau_0|^h}\frac{1}{|\tau_2 - \tau_0|^h} \frac{\sgn(\tau_1- \tau_2)}{|\tau_1 - \tau_2|^{2\Delta - h}}~,
\ee 
For $h> 2\Delta$, the term $G^0_{\chi \chi \mO}$ is much smaller than (\ref{v012}) in the IR, $\tau_{12} \gg 1$, so we can drop it in (\ref{EIG1}). Thus, (\ref{EIG1}) simplifies to (see Fig.~\ref{BS}d), 
\be \label{EIG2}
g(h)\, v(\tau_0; \tau_1, \tau_2) =  \int d\tau_3 d\tau_4\, K(\tau_1, \tau_2, \tau_3, \tau_4)\, v(\tau_0; \tau_3, \tau_4),
\ee
where $g(h) = 1$. Eqn.~\ref{EIG2} is telling us that $v(\tau_0; \tau_1, \tau_2)$ are eigenvectors of the kernel with eigenvalues $g(h)$. The dimensions $h$ of the composite operators are those $h$ for which the eigenvalue $g(h) =1$. It is helpful to think of the composite $O(N)$ invariant operators as analogous to a bound state of two fermions. 
In the more familiar context of finding bound states in quantum field theory, Fig.~\ref{BS}d is the Bethe-Salpeter equation. There one is using this equation to find the masses of the bound states. Eq.~\ref{EIG2} is the CFT analog of this, where instead of finding the masses of the bound states, one is finding the conformal dimensions $h$. 
\begin{figure}[t]
\centering
\includegraphics[width=4in]{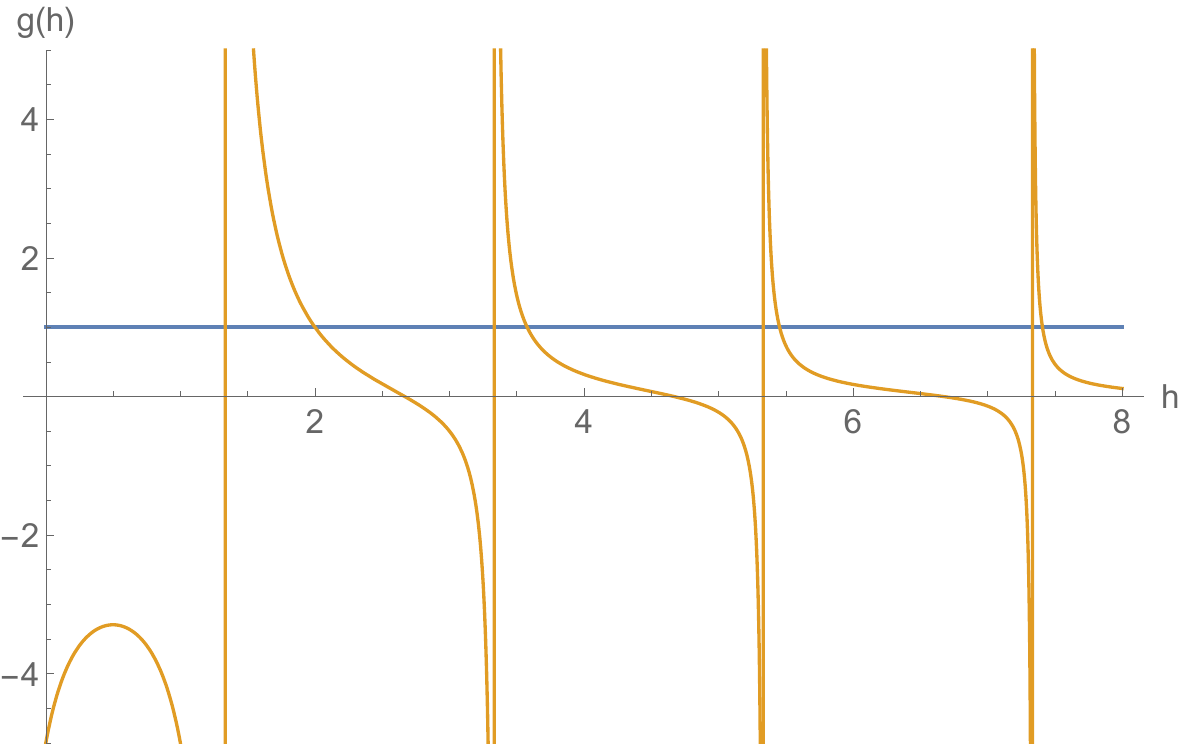}
\caption{The eigenvalues $g(h)$ (\ref{ghSYK}) of the SYK kernel (\ref{K1flav}) for $q=6$ as a function of dimension $h$. The $h$ for which $g(h)=1$ are the IR dimensions of the fermion bilinear operators (\ref{O2n1}). } \label{EigenPlotSYK}
\end{figure}

The eigenvalue $g(h)$ is independent of the choice of $\tau_0$, so for evaluating (\ref{EIG2}) one can take the eigenvectors to be, 
\be \label{val}
v (\tau_{12}) = \frac{\sgn(\tau_{12})}{|\tau_{12}|^{2 \alpha}}~,
\ee
where $2\alpha = 2\Delta - h$. By acting on (\ref{val}) with the $SL(2,R)$ generators, one gets all of the eigenvectors (\ref{v012}) \cite{PR}.  Inserting (\ref{val}) into (\ref{EIG2}) gives \cite{Kitaev}, 
\be \label{ghSYK}
g(h) = - (q-1) \frac{\psi(\Delta)}{\psi(1-\Delta)}  \frac{\psi(1- \Delta - \frac{h}{2})}{\psi(\Delta - \frac{h}{2})}~,
\ee
where $\psi(\Delta)$ was defined in (\ref{psiDelta}). A plot of $g(h)$ is given in Fig.~\ref{EigenPlotSYK}. One can see that there is a tower of $h$'s for which $g(h)=1$. For large $h$ the solutions to $g(h) = 1$ are approximately $h \approx 2\Delta+ 2n +1$. 

There are solutions to $g(h)=1$ for $h< 2\Delta$ as well. These solutions immediately follow from the $h>2 \Delta$ solutions due to the symmetry $g(h) = g(1-h)$. However, they do not correspond to dimensions of composite operators. Recall that dropping the first term in (\ref{EIG1}) was justified in the IR only for $h> 2\Delta$. (For $h<2\Delta$, it is instead justified in the UV).

Knowing the dimensions of the ``single-trace'' operators, one can say something about the bulk dual of SYK. 
The   AdS/CFT dictionary relates the dimensions of single-trace operators to the masses of bulk fields, 
\be \label{AdSCFT}
m^2 = h(h-1)~,
\ee
for AdS$_2$. So the dual of SYK has a tower of particles in the bulk, one for each solution to $g(h) = 1$ for $h>2 \Delta$. For large integer $n$, these have approximate masses $m_n^2 \approx (2\Delta + 2n+1)(2\Delta + 2n)$. 

\subsubsection{Generalized Model} \label{Gk}
\begin{figure}[t]
\centering
\subfloat[]{
\includegraphics[width=1.5in]{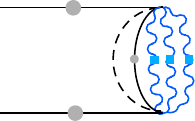} \ \ \ \ \ \ \ \ \ \ \ \ \ \ 
}
\subfloat[]{
\includegraphics[width=1.5in]{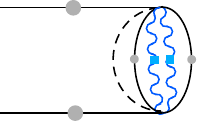}
}
\caption{The diagonal (a) and off-diagonal (b) components of the kernel (\ref{Kdiag}, \ref{Koff}) for two flavors with $q_1 = q_2=3$. The coloring scheme is the same as in Fig.~\ref{2p2t}.}\label{KerGR}
\end{figure}
We now generalize the calculation to  the  model with flavor (\ref{GFM}). The operators (\ref{O2n1}) now have a superscript $\mO_n^a$ to account for the different flavors, 
and the kernel is now a matrix in flavor space, $K^{m n}(\tau_1, \tau_2, \tau_3, \tau_4)$, where $m$ denotes the flavor of the incoming fermions on the left at times $\tau_1, \tau_2$ and $n$ denotes the flavor of the outgoing fermions on the right at times $\tau_3, \tau_4$, see Fig.~\ref{KerGR}. 
The off-diagonal component $K^{k l}$ has flavor $k$ propagators along the rails, while the rung consists of $q_k -1$ flavor $k$ propagators, $q_l-1$ flavor $l$ propagators, and $q_a$ flavor $a$ propagators for all $a \neq k,l$,  
\be
K^{k l}(\tau_1, \tau_2, \tau_3, \tau_4)= \mathfrak{b}_{k l}\, G_k(\tau_{13}) G_k(\tau_{24})\frac{1}{G_k(\tau_{34}) G_l (\tau_{34})} \prod_{a=1}^f \(G_a(\tau_{3 4})\)^{q_a},
\ee
where the combinatorial factor in front is, \footnote{The factor of $N_a^{q_a}$, for $a\neq l, k$, comes from the site index summation within the rung. For flavors $k, l$, there are only $q_k - 1, q_l-1$ propagators in the rung, so those give factors of $N_k^{q_k-1}$, $N_l^{q_l-1}$, respectively. There is then an additional factor of $N_l$ because the Feynman diagrams are built by adding the kernel to the left (see Fig.~\ref{BS}); so the $l$ index will get summed over. }
\be
\mathfrak{b}_{k l}=  -\langle J_I J_I\rangle \frac{q_k^2 q_l^2 (q_k-1)! (q_l-1)! \prod_{a\neq k,l} q_a!}{(\prod q_a!)^2}\, N_k^{q_k-1} \prod_{a\neq k} N_a^{q_a}~.
\ee
The diagonal components of the kernel are similar, but with slightly different propagator powers and combinatorial factors. Using (\ref{DisAvgGF}, \ref{GFMd}) and simplifying we get the diagonal and off-diagonal components, 
\bea \label{Kdiag}
K^{k k}(\tau_1, \tau_2, \tau_3, \tau_4) &=& - J^2 \frac{(q_k - 1)}{\kappa_k Q_k} G_k(\tau_{13}) G_k(\tau_{24})\frac{1}{G_k(\tau_{34})^{2}} \prod_{a=1}^f \(G_a(\tau_{3 4})\)^{q_a}~ \\ \label{Koff}
K^{k l}(\tau_1, \tau_2, \tau_3, \tau_4)&=& - J^2 \frac{q_l}{\kappa_k Q_k} G_k(\tau_{13}) G_k(\tau_{24})\frac{1}{G_k(\tau_{34}) G_l (\tau_{34})} \prod_{a=1}^f \(G_a(\tau_{3 4})\)^{q_a}\!,
\eea
where $k\neq l$ and $k, l \in \{1, \ldots, f\}$. If there is only one flavor, $K^{11}$ becomes (\ref{K1flav}). 

As in SYK, we must find the eigenvectors and eigenvalues of the kernel. Letting $g$ be an eigenvalue, and $v^a(\tau_{12})$  the components of an eigenvector,
\be \label{Kvv}
\sum_{b=1}^f \int d\tau_3 d\tau_4\, K^{a b}(\tau_1, \tau_2, \tau_3, \tau_4) v^b(\tau_3, \tau_4) = g\, v^a(\tau_1, \tau_2)~.
\ee
Following (\ref{val}), an ansatz for an eigenvector is,
\be \label{val2}
v^a (\tau_{12}) =c_a \frac{\sgn(\tau_{12})}{|\tau_{12}|^{2 \alpha_a}}~,
\ee
with some coefficients $c_a$. Since the eigenvector and propagators in the kernel only depend on time differences, (\ref{Kvv}) factorizes nicely under a Fourier transform, 
\be \label{EigFT}
\frac{J^2}{\kappa_a Q_a}\mF(G_a)^2\Big[ (q_a-1)  \mF\( \frac{\prod G_c^{q_c}}{G_a^2} v^a\) + \sum_{b\neq a} q_b \mF\(\frac{\prod G_c^{q_c}}{G_a G_b} v^b\)
\Big] = g \mF(v^a)~,
\ee
where the first term on the left is from the diagonal term in the kernel (\ref{Kdiag}) and the second term is from the off-diagonal terms (\ref{Koff}). Inserting the propagator (\ref{Gkto}) and evaluating gives, 
\begin{multline} \label{EigFT2}
\frac{ (\prod b_k^{q_k})}{\kappa_a Q_a} \, \psi(\Delta_a)^2\[ c_a (q_a - 1)\, \psi(1- 2\Delta_a + \alpha_a)  \right.+\\
 \left. \sum_{b\neq a}\(\frac{J}{|\omega|}\)^{2 \Delta_b - 2\Delta_a} |\omega|^{2 (\alpha_b - \alpha_a)}\frac{b_a}{b_b}c_b q_b\, \psi(1- \Delta_a - \Delta_b + \alpha_b)\] = g\,c_a \psi(\alpha_a)~.
\end{multline}
In order to eliminate the dependance on $\omega$, we must choose, 
\be \label{aba}
\alpha_b = \alpha_a + \Delta_b - \Delta_a~.
\ee
In SYK we know that (\ref{val}) is a special case of (\ref{v012}) with $2\alpha = 2\Delta - h$. Similarly here, we let
\be
2\alpha_a = 2\Delta_a - h~,
\ee
which is consistent with (\ref{aba}). Thus, (\ref{EigFT2}) becomes an eigenvector equation for the matrix $\tilde{K}$, 
\be
\tilde{K}\, \vec{c} = g\, \vec{c}~,
\ee
where the diagonal and off diagonal components of $\tilde{K}$ are, 
\bea \label{Ktilde2}
\tilde{K}^{aa} &=& (q_a - 1)\, \rho_a (h)\\ \nonumber
\tilde{K}^{ab} &=& q_b\,  \frac{b_a J^{-2\Delta_a}}{b_b J^{-2\Delta_b}}\, \rho_a(h)~,
\eea
where
\be \label{rhoaa}
\rho_a(h) = -\frac{\psi(\Delta_a)}{\psi(1-\Delta_a)}  \frac{\psi(1- \Delta_a - \frac{h}{2})}{\psi(\Delta_a - \frac{h}{2})}~.
\ee
In getting from (\ref{EigFT2}) to (\ref{Ktilde2}) we made use of the product of normalizations of the propagators $\prod b_a^{q_a}$ given in  (\ref{GenNorm1}). If there is one flavor, $\tilde{K}^{11}$ reduces to (\ref{ghSYK}). 

The next step is to find all $h$ for which there is an eigenvalue $g$ of  $\tilde{K}$ (\ref{Ktilde2}) that equals $1$. This is in principle straightforward: for any $q_a$, $\kappa_a$ in (\ref{GFM}) ones solves  (\ref{DimSumG}, \ref{GenNorm2}) to find the IR dimensions $\Delta_a$ of the fermions, then for fixed $h$ one  finds the $f$ eigenvalues of $\tilde{K}$, and then for each of those eigenvalues solves for all $h$ such that the eigenvalue is equal to one. Aside from some special cases, we can not write a general and explicit answer for the $h$'s. However, it is easy to see that there will always be a dimension $2$ operator in the spectrum. For $h=2$ (\ref{rhoaa}) simplifies to,
\be
\rho_a(h=2) =  \frac{\Delta_a}{1-\Delta_a}~.
\ee
Inserting this into (\ref{Ktilde2}), one can easily verify that the following vector
\be
v^a = \Delta_a b_a J^{-2\Delta_a}~,
\ee
is an eigenvector of $\tilde{K}$ with eigenvalue one. Verifying this requires using $\sum q_a \Delta_a = 1$, and nothing else. Perhaps surprisingly, it is not even required that the $\Delta_a$ are actual dimensions: one does not need to impose (\ref{GenNorm2}). The dimension-two operator is important: it leads to the breaking of conformal invariance and to maximal chaos; we will comment more on it in the next section. Another universal feature (for any number of flavors greater than one) is the seeming presence of a dimension-one operator. Inserting $\rho(h=1) = -1$  into (\ref{Ktilde2}), one can verify that there are $f-1$ eigenvectors of $\tilde{K}$ that have eigenvalue one. For any $k \in \{2,\ldots, f\}$, such an eigenvector has two nonzero components,
\be
v_1 = - b_1 J^{2\Delta_k} q_k~, \ \ \ \ \ \ v_k = b_k J^{2\Delta_1} q_1~.
\ee
In fact, verifying this requires no assumptions on $\Delta_a$. The presence of these dimension-one operators suggests a symmetry. In fact, this symmetry is simple to see from the effective action (\ref{Zeff}).~\footnote{We thank J.~Maldacena for recognizing this.} One can rescale $\tilde{G}_1(\tau_1,\tau_2) \rightarrow f(\tau_1) f(\tau_2) \tilde{G}_1(\tau_1, \tau_2)$ and $\tilde{G}_a(\tau_1,\tau_2) \rightarrow [f(\tau_1) f(\tau_2)]^{-\frac{q_1}{q_a}} \tilde{G}_a(\tau_1, \tau_2)$, for any $a\neq 1$, while leaving the IR limit of (\ref{Zeff}) invariant.

\subsubsection{Equal $q_a$, $\kappa_a$} \label{equalD}
\begin{figure}[t]
\centering
\includegraphics[width=4in]{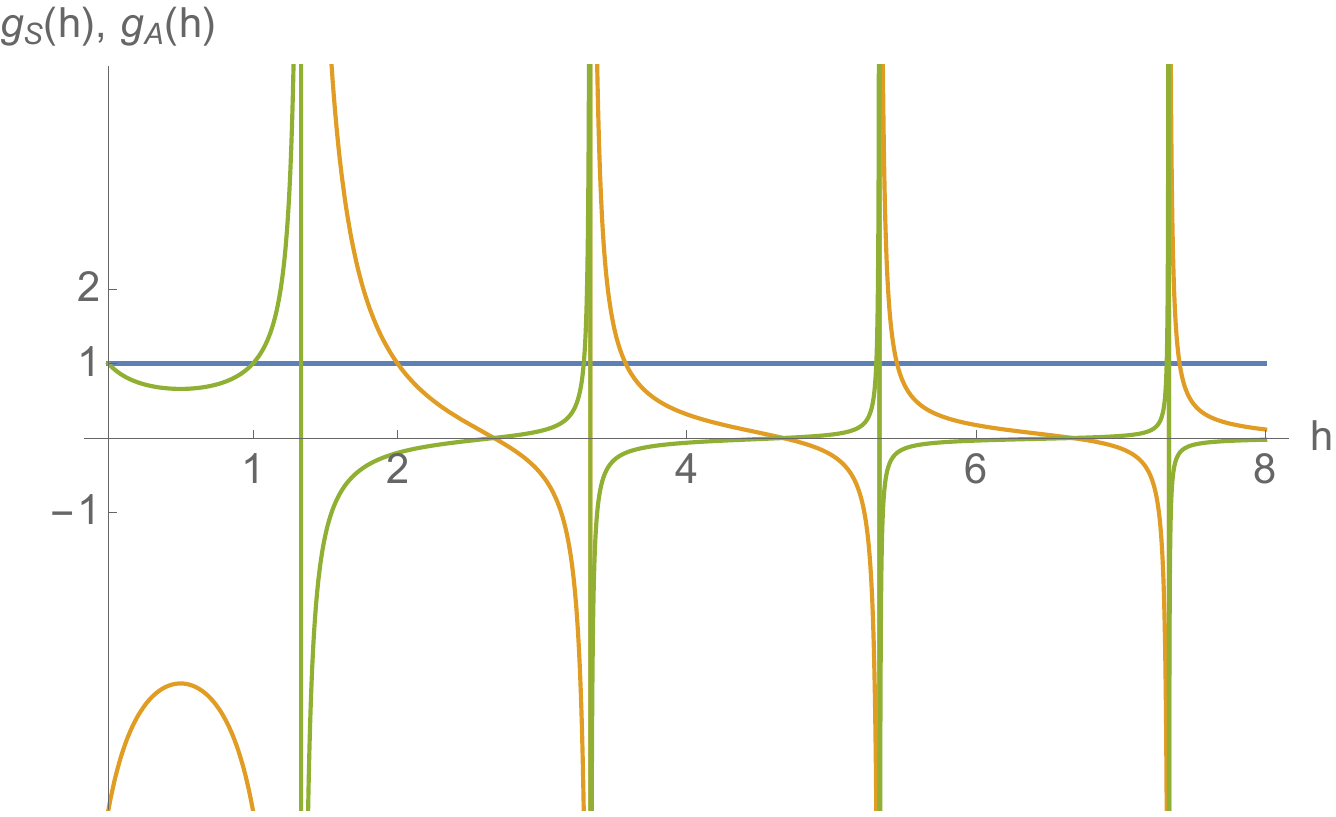}
\caption{The two families $g_S(h), g_A(h)$ of eigenvalues (\ref{gkh}) of the kernel for the two-flavor generalized model (\ref{Kdiag}, \ref{Koff}) with $q_1=q_2=3$, $\kappa_1=\kappa_2=1/2$. The $g_S(h)$ family is the same as SYK with $q=6$, see Fig.~\ref{EigenPlotSYK}. The IR dimensions of the bilinear fermion singlets are those $h$ for which either $g_S(h)=1$ or $g_A(h)=1$.  } \label{GRspec}
\end{figure}
A simple and instructive case is when all the $q_a$  are equal to some $q$, and all the $\kappa_a$ are equal, for all flavors $a$. The dimensions $\Delta_a$ are then, by symmetry, all equal to $\Delta = \frac{1}{f q}$. The matrix $\tilde{K}$ in (\ref{Ktilde2}) factorizes, 
\be \label{mK}
\tilde{K} = \rho(h) \mK~, \ \ \ \ \ \text{where   }\, \,\, \,  \mK^{aa} = (q-1), \ \ \ \mK^{ab} = q~,
\ee
where $\rho(h)$ is given by (\ref{rhoaa}) and is independent of the flavor. The eigenvalues of $\tilde{K}$ are thus, 
\be \label{gkh}
g_k(h) = \rho(h) \sigma_k,
\ee
where $\sigma_k$ are the $f$ eigenvalues of $\mK$. The matrix $\mK$ has a  symmetric eigenvector  $(1,1, \ldots, 1)$ with eigenvalue $\sigma = f q-1$, as well as $f-1$ antisymmetric eigenvectors: $(1,-1,0,\ldots, 0)$, $(1,0,-1, 0, \ldots, 0)$, $\ldots, (1,0, \ldots, 0, -1)$, all with the  same eigenvalue $\sigma = -1$. Setting $g_k(h)$ equal to $1$ gives the dimensions $h$. The eigenvalue  $\sigma = f q-1$ leads to the same tower of dimensions as SYK with an $f q$ body interaction, while the eigenvalue $\sigma = -1$ gives an additional and new tower of operators, see Fig.~\ref{GRspec}. The origin of the new towers is due to the more refined symmetry of the generalized model as compared to SYK: a product of $f$ $O(N)$'s instead of $O(N f)$. 

An alternative way to think about the generalized model (\ref{GFM}) for this case is that instead of having $f$ flavors of fermions with $N_1=N_2= \ldots = N_f$ sites for each, there is one flavor with $N_1 f$ sites. In other words, in the Hamiltonian (\ref{GFM}), 
\be
\sum_I J_I (\chi_{i^{(1)}_1}^1 \cdots \chi_{i^{(1)}_{q}}^1) (\chi_{i^{(2)}_1}^2 \cdots \chi_{i^{(2)}_{q}}^2) \cdots (\chi_{i^{(f)}_1}^f \cdots \chi_{i^{(f)}_{q}}^1) ~,
\ee
where $I = i^{(1)}_1, \ldots, i^{(1)}_q, \ldots, i^{(f)}_1, \ldots, i^{(f)}_q$, one makes the identification $i^{(p)}_k = f n_k + p-1$, where $n_k$ ranges from $1$ to $N_1$,  and gets rid of the flavor index on the fermions. There are now $f N_1$ sites; however, this is not the same as SYK with a $q f$ body interaction, since the interactions are not all-to-all, being restricted to occur between  particular $q f$ sets of sites.

\subsection{Four-Point Function}

Having found the dimensions  of the bilinear singlet operators $\mO_n^a$, the next step is to compute their OPE coefficient. The OPE between two fermions will include all the $\mO_n^a$ and their descendants, and will take the form, 
\be \label{OPEcc}
\frac{1}{N_a}\sum_{i=1}^{N_a} \chi_i^a(\tau_1) \chi_i^a(\tau_2) = \sum_{n, b} c_n^{a, b}\, \mathcal{C}_n(\tau_{12}, \partial_{\tau_1})\mO_n^b (\tau_1)~,
\ee
where $c_n^{a, b}$ are the OPE coefficients and $\mC_n(\tau_{12}, \partial_{\tau_1}) = 1 + \ldots$ is fixed by conformal invariance. The OPE coefficient  can be extracted by computing the three-point function between the two fermions  and $\mO$, which in Sec.~\ref{DimComp} was labelled as $v(\tau_0; \tau_1, \tau_2)$ and satisfied Eq.~\ref{EIG1}. Thinking of $K$ as a matrix with indices $(\tau_1, \tau_2)$, $(\tau_3, \tau_4)$, the formal solution of (\ref{EIG1}) is, 
\be \label{v1mK}
v^a(\tau_0; \tau_1, \tau_2) = \frac{1}{1- K} G_{\chi^a \chi^a \mO}^0~,
\ee
where we have generalized (\ref{EIG1}) to account for multiple flavors. Notice that when we computed the dimensions in Sec.~\ref{DimComp}, we were allowed to drop the $G_{\chi \chi\mO}^0$ term in (\ref{EIG1}), arguing it was unimportant in the IR. However, for finding the OPE coefficients one is interested in the UV, $\tau_{12} \ll 1$, so this term is essential. 

We will also be interested in the four-point function. Defining the bilocal, 
\be \label{bilocal}
g_a(\tau_1, \tau_2) \equiv \frac{1}{N_a}  \sum_{i=1}^{N_a} \chi_i^a(\tau_1) \chi_i^a(\tau_2)~,
\ee
and proceeding formally, one can perform a double OPE expansion on the four-point function,~\footnote{Here $n$ ranges over the positive integers and is labeling the number of derivatives in the composite operator, see (\ref{O2n1}). The index $e$ is labeling the different flavors. In writing (\ref{4pt2F0}) we have assumed, as will generically be the case, that there are no degeneracies in the dimensions  $h_{n, e}$ of the composites.}
\be \label{4pt2F0}
\langle g_a(\tau_1, \tau_2) g_b(\tau_3, \tau_4) \rangle = \frac{1}{N_a N_b} \sum_{n, e} c_n^{a, e} c_n^{b, e}\,  \mC_n(\tau_{12}, \partial_{\tau_1}) \mC_n(\tau_{34}, \partial_{\tau_3}) \frac{1}{|\tau_{13}|^{2 h_{n,e}}}.
\ee
The right-hand side is a sum of conformal blocks, given by hypergeometric functions of the conformally invariant cross ratio,
\be \label{4pt2F1}
\langle g_a(\tau_1, \tau_2) g_b(\tau_3, \tau_4) \rangle = G_a(\tau_{12}) G_b(\tau_{34}) \sum_{n, e} c_n^{a, e} c_n^{b, e}\, x^{h_{n, e}}\, {_2} F_1(h_{n, e}, h_{n, e}, 2 h_{n, e}, x)~, \ \ \ \ \ \ x= \frac{\tau_{12} \tau_{34}}{\tau_{13} \tau_{24}}~.
\ee
This is similar to two-dimensional CFTs, except here we have one cross-ratio instead of two. 
\begin{figure}[t]
\centering
\subfloat[]{
\includegraphics[width=1.3in]{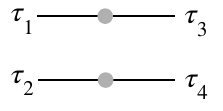} \ \ \ \ \ \ \ \ \ \ \ \ \ \ 
}
\subfloat[]{
\includegraphics[width=1.3in]{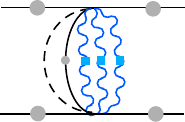}\ \ \ \ \ \ \ \ \ \ \ \ \ \ 
}
\subfloat[]{
\includegraphics[width=1.3in]{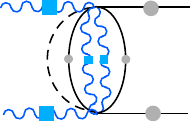}
}
\caption{The $1/N$ piece of the four-point function (\ref{gagb}) consists of a disconected piece, $\mF_0$,  which is diagonal in flavor space, with the first term in (\ref{F0ab}) shown in (a). The ladder diagrams are formed by acting with the kernel (\ref{Kdiag}, \ref{Koff}) on the left of $\mF_0$ to add rungs. Adding one rung gives diagrams such as those in (b) and (c). }\label{4ptGen}
\end{figure}
 At large $N$, the leading and first subleading in $1/N$ pieces of the four-point function are, 
 \be \label{gagb}
 \langle g_a(\tau_1, \tau_2) g_b(\tau_3, \tau_4) \rangle = G_a(\tau_{12}) G_b(\tau_{34}) + \frac{1}{N} \mF^{a b} (\tau_1, \tau_2, \tau_3, \tau_4)~.
 \ee
The $1/N$ piece of the  four-point function is found by summing ladder diagrams \cite{Kitaev}. What we have is a slight generalization of what occurs in SYK, as the four-point function is now a matrix in flavor space. Starting with
\be \label{F0ab}
\mF_0^{a b}(\tau_1, \tau_2, \tau_3, \tau_4) = \delta^{ab}\(- G_a(\tau_{13}) G_a(\tau_{24}) + G_a(\tau_{14}) G_a(\tau_{23})\) ~,
\ee
one uses the kernel  (\ref{Kdiag}, \ref{Koff}) to add rungs to the ladder. Summing all the ladder diagrams,
\be \label{Fab}
\mF^{ab} (\tau_1, \tau_2, \tau_3, \tau_4) =\frac{1}{\kappa_b} \(\frac{1}{1- K} \mF_0\)^{ab}~.
\ee

\subsubsection*{SYK}
The technical challenge in evaluating (\ref{Fab}) explicitly comes from inverting $1-K$. 
Recall the procedure used in SYK. One  first finds a complete basis of eigenvectors of the kernel. This turns out to be given by (\ref{v012}) with $h$ ranging over even positive integers $h = 2, 4, 6, \ldots$, as well as $h=1/2 + i s$ where $s>0$ \cite{KitaevUn, PR, MS}. One then projects (\ref{Fab}) onto this basis and performs the sum/integral over the discrete and continuous tower of $h$'s to find (\ref{4pt2F1}) with OPE coefficients $c_n$ \cite{MS}, 
\be \label{cnSYK}
(c_n)^2 =  \alpha_0(q)\frac{(h_n - 1/2)}{\pi \tan(\pi h_n/2)} \frac{\Gamma(h_n)^2}{\Gamma(2 h_n)} \frac{1}{g'(h_n)} ~,  \ \ \ \ \  \text{where}\ \ \alpha_0(q) = \frac{2\pi q}{(q-1) (q-2) \tan \frac{\pi}{q}}~,
\ee
where $g(h)$ is given by (\ref{ghSYK}) and $h_n$ are the solutions of $g(h_n) =1$.  

Eq.~\ref{cnSYK} is for $h_n>2$. There is an additional complication that occurs for the $h=2$ block. One can notice that $g(h=2) = 1$, and since $h=2$ is part of the basis of eigenvectors used to invert $1-K$, this causes the four-point function to diverge in the conformal limit. The $h=2$ block must therefore be treated outside the conformal limit. Moving slightly away from the IR, the eigenvalue $g(h=2)$ gets slightly shifted away from $1$, and so the $h=2$ block gives a finite but large, and non-conformal, contribution to the four-point function. Since its prefactor is dominant, its growth controls the behavior of the finite temperature out-of-time-order four-point function used to probe chaos \cite{LO, KitaevNov}. The growth of the $h=2$ block occurs with a Lyapunov exponent $2\pi T$ that saturates the chaos bound of \cite{MSS}.~\footnote{At strong coupling, the SYK Lyapunov exponent only depends on the temperature $T$. At weak coupling, the SYK Lyapunov exponent scales with the coupling $J$ \cite{Kitaev}.}  In Sec.~\ref{Gk} we found that the generalized model always contains a dimension-two operator; assuming its OPE coefficient doesn't vanish, this implies that in the IR the generalized model, like SYK,  both breaks conformal invariance and is maximally chaotic. 
\newline

To compute the four-point function (\ref{Fab}) for the generalized model with generic $q_a$ and $\kappa_a$, one would need to repeat the procedure used for SYK, accounting for the additional complexity of having flavor. However, in the case that all the $q_a$ are equal and all the $\kappa_a$ are equal, it is simple to find the four-point function, and this is case we focus on. 

\subsubsection{Equal $q_a$, $\kappa_a$}
If all the $q_a$ are equal to $q$, and all the $\kappa_a = 1/f$, then the kernel matrix (\ref{Kdiag}, \ref{Koff}) factorizes into a flavor-space matrix and a function of times,
\be
K(\tau_1, \tau_2, \tau_3, \tau_4) = \mK\, \mathit{k}(\tau_1, \tau_2, \tau_3, \tau_4)~, \ \ \ \ \ \ \  \ \ \ \mathit{k}(\tau_1, \tau_2, \tau_3, \tau_4)= - J^2\frac{f}{q^{f-1}}G(\tau_{13}) G(\tau_{24}) G(\tau_{34})^{ f q -2}~,
\ee
where $\mK$ was defined in  (\ref{mK}). By symmetry, the two-point functions are flavor-independent, $G_a(\tau) \equiv G(\tau)$. For concreteness, let us focus on the case of two flavors, $f=2$. In Sec.~\ref{equalD} we diagonalized $\mK$, finding a symmetric eigenvector: $(1,1)$, with eigenvalue $\sigma_S = 2 q-1$, and an antisymmetric eigenvector: $(1,-1)$, with eigenvalue $\sigma_A = -1$. Forming a matrix of the eigenvectors, 
\be
O = \frac{1}{\sqrt{2}} \left(\begin{array}{cc}
1 & 1 \\
1 & - 1\end{array}\right)~,
\ee
we diagonalize (\ref{Fab}) in flavor space, forming $O^T \mF O$, to find, 
\bea
\mF^{11} &=& \frac{1}{1 - \sigma_S \mathit{k}} F_0 +  \frac{1}{1 - \sigma_A \mathit{k}} F_0~,\\
\mF^{12} &=& \frac{1}{1 - \sigma_S \mathit{k}} F_0 -  \frac{1}{1 - \sigma_A \mathit{k}} F_0~,\\
\mF^{12} &=& \mF^{21}~, \ \ \ \ \ \ \ \mF^{11} = \mF^{22}~,
\eea
where  $ F_0 = - G(\tau_{13}) G(\tau_{24}) + G(\tau_{14}) G(\tau_{23})$ is the diagonal component of $\mF_0$ in (\ref{F0ab}). Both of the terms appearing in $\mF^{ab}$ are similar to what occurs in SYK, 
so we can write the answer, 
\be
 \frac{1}{1 - \sigma_{S/A} \mathit{k}} F_0 = G(\tau_{12}) G(\tau_{34}) \sum_n (c_n^{S/A})^2\,  x^{h_n}\, {_2} F_1(h_n, h_n, 2 h_n, x)
\ee
where,
\bea \label{cs}
(c_n^S)^2 &=& \alpha_0(2 q) \frac{(h_n - 1/2)}{\pi \tan(\pi h_n/2)} \frac{\Gamma(h_n)^2}{\Gamma(2 h_n)} \frac{1}{(2 q-1) \rho'(h_n)}~,   \ \text{for} \ \ \ \ (2q-1) \rho(h_n) = 1~,\\ \label{ca}
(c_n^A)^2 &=& \alpha_0(2 q) \frac{(h_n - 1/2)}{\pi \tan(\pi h_n/2)} \frac{\Gamma(h_n)^2}{\Gamma(2 h_n)} \frac{(2q-1)}{ \rho'(h_n)}~,  \   \ \ \ \ \  \ \ \ \text{for} \ \ \ \ \ \ \  \ \   -\rho(h_n) = 1~.
\eea
Here $\alpha_0(q)$ is given by (\ref{cnSYK}) and $\rho(h)$ is given by (\ref{rhoaa}) with fermion dimension $\Delta = 1/2 q$. To be clear, the $h_n$ appearing in (\ref{cs}) and (\ref{ca}) are the solutions of $ (2q-1) \rho(h_n) = 1$ and $ -\rho(h_n) = 1$, respectively. Recall that we found in Sec.~\ref{equalD} that with two flavors (with $q_1=q_2=q$, $\kappa_1 = \kappa_2$),  the spectrum of bilinear composite operators contains two towers: a tower that matches the $ 2 q$ body SYK tower, and a new tower, see Fig.~\ref{GRspec}. The OPE coefficients $c_n^S$ are for the $ 2 q$ body SYK tower. Note that  (\ref{cs}) is for $h_n>2$; as discussed before, the contribution of the $h=2$ block diverges in the conformal limit. 

The OPE coefficients $c_n^A$ are for the new tower. Notice that this vanishes for the $h=1$ operator. 
The OPE coefficients $c_n^1$ and $c_n^2$, in terms of (\ref{cs}, \ref{ca}), are given by, 
\bea \label{cn1S}
c_n^1 &=& c_n^2~, \ \ \ \ (c_n^1)^2 = (c_n^S)^2 \  \ \ \  \ \ \text{for} \ \ (2q-1) \rho(h_n) = 1~, \\ \label{cn1A}
c_n^1 &=& - c_n^2, \ \ \  (c_n^1)^2 = (c_n^A)^2 \ \ \ \ \   \text{for} \ \ \ \ \ \  \ \ \    -\rho(h_n) = 1~,
\eea
where, for simplicity of presentation, rather than writing $c_n^{a, b}$,  we have explicitly separated the two towers.~\footnote{The $c_n^1$ in (\ref{cn1S}) are the OPE coefficients for two fermions of flavor $1$ going into the sum of $\mO_n^1$ and $\mO_n^2$ (each of which is given by (\ref{O2n1}) for the corresponding flavor), while the $c_n^1$ in (\ref{cn1A}) are the OPE coefficients for two fermions of flavor $1$ going into the difference between $\mO_n^1$ and $\mO_n^2$ . Analogously for the fermions of flavor $2$ and the $c_n^2$.}
A more intuitive way to think about this four-point function is to define the symmetric and antisymmetric combinations of the bilocals (\ref{bilocal}),  
\bea \label{gsa}
g_S(\tau_1, \tau_2) &=& \frac{1}{2} \( g_1(\tau_1, \tau_2) + g_2(\tau_1, \tau_2) \) \\ \nonumber
g_A(\tau_1, \tau_2) &=& \frac{1}{2} \( g_1(\tau_1, \tau_2) - g_2(\tau_1, \tau_2) \)~.
\eea
The symmetric correlator probes only the SYK tower, 
\be \label{gsgs}
\langle g_S(\tau_1, \tau_2) g_S(\tau_3, \tau_4) \rangle =  G(\tau_{12}) G(\tau_{34})+  \frac{1}{N}  \frac{1}{1 - \sigma_S \mathit{k}} F_0
\ee
and matches the SYK $ 2 q$ body four-point function. The antisymmetric correlator, 
\be \label{gaga}
\langle g_A(\tau_1, \tau_2) g_A(\tau_3, \tau_4) \rangle = \frac{1}{N}  \frac{1}{1 - \sigma_A \mathit{k}} F_0~,
\ee
probes only the new tower. One can also reproduce (\ref{gsgs}, \ref{gaga}) from the path integral picture, see Appendix~\ref{AppFluc}.

\section{Discussion} \label{sec:Dis}
The SY model \cite{SY} involves all-to-all interactions between spins in some representation of $SU(M)$, $H = \sum_{ i, j = 1}^N \sum_{\mu, \nu =1}^M J_{i j} S_{i\, \nu}^{\mu}S_{j\, \mu}^{\nu}$, with Gaussian-random couplings $J_{i j}$. 
Writing  the spins as products of two fermions, this becomes a four-fermion interaction. One of the key realizations of \cite{SY} was that the model is solvable in the double scaling limit, $N\rightarrow \infty$, $M\rightarrow \infty$, $M/N \rightarrow 0$. It was recognized in \cite{Kitaev} that a simpler model is one that avoids spins altogether and goes directly to the fermions, $H= \sum J_{ i j k l}\,  \chi_i \chi_j \chi_k \chi_l$. There is then  a clear generalization to a model with a $q$-index coupling and a $q$-body interaction \cite{Kitaev}. 

In this paper, we have made another straightforward generalization, involving $f$ flavors of fermions with $N_a$ sites for each flavor and a $\sum_{a=1}^f q_a$ body interaction. Perhaps surprisingly, the model has an infrared fixed point for most choices of parameters $N_a, q_a$. We found a set of equations determining the dimensions of the fermions in the infrared, as well as the matrix determining the infrared dimensions of the bilinear singlet operators that are invariant under the global $O(N_1) \times O(N_2) \times \cdots \times O(N_f)$ symmetry.

It was recognized in \cite{MS} that the SYK model simplifies in the limit $q \gg 1$. Here we pointed out that in the $q \gg 1$ limit, only a particular subset of Feynman diagrams need to be summed. For any even $q\geq 4$, the SYK model has qualitatively similar properties. In the generalized model introduced in this paper, there are more parameters to vary, and one may wonder if there are corners of parameter space which either lead to simplifications or qualitative differences. 

We have only begun exploring the parameter space, focusing on the symmetric case of an equal number of sites for each flavor, as well as interaction orders $q_a$ that are independent of the flavor. The main qualitative difference, as compared to SYK with a $ q f$ body interaction, is more  singlet operators resulting from a symmetry that is a subgroup of the $O(N)$ symmetry of SYK. One  feature we found, that holds for any choice of parameters, is the presence of a dimension-two bilinear singlet operator in the infrared. Another was the presence of a dimension-one operator; however, in the symmetric case considered, its OPE coefficient vanished. It would be good to better understand this operator. 

Nontrivial and solvable models are both rare and valuable. It is now clear that the class of SYK-like models is much larger than just the SY model. Just how large this class is, if there are further generalizations, and the precise characterization of  the Feynman diagrams, at each order in $1/N$, are all still open problems. We may hope that exploring this structure will provide guidance towards understanding the dual string theory, if there is one.

\bigskip

\section*{Acknowledgements}\noindent We thank  D.~Anninos,  T.~Anous, D.~Gaiotto, A.~Kitaev, G.~Korchemsky,  J.~Maldacena, Y.~Nakayama, N.~Nekrasov, J.~Polchinski, B.~Shraiman, and  E.~Silverstein for helpful discussions.  This work was supported by NSF grant 1125915.

\appendix
\section{Effective Action} \label{free}
In this appendix we compute the free energy (equivalently, the effective action) for the generalized model (\ref{GFM}). The calculation is analogous to the one for SYK \cite{Kitaev}. 

Employing the replica trick, instead of computing the disorder average of the logarithm of the partition function, one instead computes the disorder average of $M$ copies of the system. Starting with (\ref{GFM}), this is given by,
\bea \label{Zm1} \nonumber
\overline{Z^M} = \int D \chi_i^{a, \al}\, D J_{I}\,\,  P[J_I]\,  \exp\[ - \sum_{\al=1}^M \int d\tau\(\frac{1}{2}\sum_{a=1}^f \sum_{i_a=1}^{N_a}  \chi_{i_a}^{a, \al} \partial_{\tau}\, \chi_{i_a}^{a, \al} \right.\right.\\
\left. \left.+ \frac{(i)^{\frac{\mathrm{q} }{2}}}{\prod_{a=1}^f q_a!} \sum_I J_{I}( \chi_{i_1}^{1,\alpha} \cdots \chi_{i_{q_1}}^{1,\alpha})\cdots (\chi_{j_1}^{f,\alpha} \cdots \chi_{j_{q_f}}^{f,\alpha})\)\]
\eea
where $\alpha$ is the replica index, $\al\in\{1,\ldots,M\}$, $a$ is the flavor, $a\in\{1,\ldots, f\}$, $i_a$ is the site index, $i_a \in\{1, \ldots, N_a\}$, and $I$ is a collective site index,  $I = i_1,\ldots, i_{q_1}, \ldots, j_{1},\ldots, j_{q_f}$, and $P[J_I]$ is the probability distribution for the $J_I$ (\ref{PJI}).   Doing the Gaussian integral over the disorder, (\ref{Zm1}) becomes, 
\begin{multline} \label{Zm2}
\overline{Z^M} = \int D \chi_{i}^{a, \alpha}\, \exp\( - \sum_{\al=1}^M\sum_{a=1}^f \sum_{i_a=1}^{N_a} \frac{1}{2} \int\, d\tau\, \chi_{i_a}^{a, \al} \partial_{\tau}\, \chi_{i_a}^{a, \al}\right. \\
\left.+ \frac{J^2\, N}{2 (\prod_a q_a)} \sum_{\alpha, \beta} \int d\tau_1 d\tau_2\, \prod_a \(\sum_{i_a=1}^{N_a}\frac{1}{N_a} \chi_{i_a}^{a, \alpha}(\tau_1) \chi_{i_a}^{a, \beta}(\tau_2)\)^{q_a} \)~.
\end{multline}
Having done the disorder average, we see that there is a $O(N_1)\times O(N_2) \times \cdots \times O(N_f)$ symmetry. We thus introduce the collective fields,
\be
\tG_a^{\al \beta}(\tau_1, \tau_2)= \frac{1}{N_a} \sum_{i_a=1}^{N_a} \chi_{i_a}^{a, \al}(\tau_1) \chi_{i_a}^{a, \beta}(\tau_2)
\ee
\vspace{-.1em}
by inserting delta functions,
\begin{multline}
\delta\(\tG_a^{\al \beta}(\tau_1, \tau_2)- \frac{1}{N_a} \sum_{i_a=1}^{N_a} \chi_{i_a}^{a, \al}(\tau_1) \chi_{i_a}^{a, \beta}(\tau_2)\) \\
\propto \int d \tS_a^{\al \beta}(\tau_1, \tau_2)\, \exp\( - \frac{N_a }{2}\, \tS_a^{\al \beta}(\tau_1, \tau_2) \(\tG_a^{\al \beta}(\tau_1, \tau_2)- \frac{1}{N_a} \sum_{i_a=1}^{N_a} \chi_{i_a}^{a, \al}(\tau_1) \chi_{i_a}^{a, \beta}(\tau_2)\)\)~,
\end{multline}
where $ \tS_a^{\al \beta}(\tau_1, \tau_2)$ acts as a Lagrange multiplier. We insert into (\ref{Zm2}) such a delta function for each replica index pair $\al, \beta$ and each flavor $a$. This gives, 
\begin{multline}
\overline{Z^M} = \int D\chi_{i_a}^{a \al}\, D\tS_a^{\al \beta}\, D\tG_a^{\al \beta}\, \\
\exp\left(- \sum_{\al,\beta=1}^M \sum_{a=1}^f \sum_{i_a=1}^{N_a}\, \frac{1}{2}\int d\tau_1 d\tau_2\, \chi_{i_a}^{a, \al}(\tau_1) \(\delta_{\al \beta} \delta(\tau_{12})\, \partial_{\tau} - \tS_a^{\al \beta}(\tau_1, \tau_2)\)\chi_{i_a}^{a, \beta}(\tau_2)  \right. \\
\left. -\frac{1}{2} \sum_{\al, \beta=1}^M\int d\tau_1 d\tau_2\, \(\sum_{a=1}^f N_a \tS_a^{\al \beta}(\tau_1, \tau_2) \tG_a^{\al \beta}(\tau_1, \tau_2) - \frac{J^2 N}{\prod q_a} \prod_a \(G_a^{\al \beta} (\tau_1, \tau_2)\)^{q_a}\)\)~.
\end{multline}
Integrating out the fermions gives
\be \label{ZZ}
\overline{Z^M} = \int D\tS_a^{\al \beta}\, D\tG_a^{\al \beta} \exp\( - \sum_{\al, \beta =1}^M  S_{eff}^2\)
\ee
where 
\begin{multline}
S_{eff}^2 = - \frac{1}{2}\sum_{a=1}^f N_a\,\log \det \(\delta_{\al \beta}\partial_{\tau} - \tS_a^{\al \beta}\) \\
+ \frac{1}{2} \int d\tau_1 d\tau_2 \( \sum_{a=1}^f N_a\, \tS_a^{\al \beta}(\tau_1, \tau_2) \tG_a^{\al \beta}(\tau_1, \tau_2) - \frac{J^2 N}{\prod_a q_a} \prod_a \(\tG_a^{\al \beta}(\tau_1, \tau_2)\)^{q_a}\)
\end{multline}
As is standard in studies of SYK, one assumes a replica symmetric saddle point, $\tG_a^{\al \beta}(\tau_1, \tau_2)=\delta^{\alpha \beta} \tG_a (\tau_1, \tau_2)$, and so (\ref{ZZ}) becomes $\overline{Z^M} = \int D\tS_a\, D\tG_a \exp\( - M S_{eff}\)~$
where,
\begin{multline} \label{SeffApp}
S_{eff} = - \frac{1}{2} \sum_{a=1}^f N_a\, \log \det \(\partial_{\tau} - \tS_a\) \\
+ \frac{1}{2} \int d\tau_1 d\tau_2 \(\[\sum_{a=1}^f N_a\, \tS_a (\tau_1, \tau_2) \tG_a(\tau_1, \tau_2)\] - \frac{J^2 N}{\prod_a q_a} \prod_a \(\tG_a(\tau_1, \tau_2)\)^{q_a}\)~.
\end{multline}
If there is only one flavor, we recover the SYK action \cite{Kitaev},
\be \label{SEFF}
S_{eff}/N = - \frac{1}{2} \log \det \(\partial_{\tau} - \tS\) + \frac{1}{2} \int d\tau_1 d\tau_2 \( \tS(\tau_1, \tau_2) \tG(\tau_1, \tau_2) - \frac{J^2}{q} \tG(\tau_1, \tau_2)^q\)~.
\ee
\subsection{Fluctuations} \label{AppFluc}
For SYK, one can expand (\ref{SEFF}) about the saddle $\tG = G + |G|^{\frac{2-q}{2}} g$ and $\tS = \Sigma + |G|^{\frac{q-2}{2}}\sigma$, keeping terms up to second order, and then integrating out $\sigma$  to get \cite{MS}, 
\be \label{sqwe}
\!\!\frac{S_{eff}}{N} = \frac{1}{4} \int d\tau_1\ldots d\tau_4\, g(\tau_1, \tau_2) K_c^{-1} (\tau_1, \ldots, \tau_4)  g(\tau_3, \tau_4) -\frac{J^2 (q-1)}{4}\int d\tau_1 d\tau_2\, g(\tau_1, \tau_2)^2 
\ee
where $K_c^{-1}$ is the inverse of $K_c$, thought of as a matrix with indices $(\tau_1, \tau_2)$, $(\tau_3, \tau_4)$, and given by, 
\be
K_c(\tau_1, \ldots, \tau_4) =  - |G(\tau_1, \tau_2)|^{\frac{q-2}{2}}G(\tau_1, \tau_3) G(\tau_2, \tau_4) |G(\tau_3, \tau_4)|^{\frac{q-2}{2}}~.
\ee
We can write (\ref{sqwe}) in the shorthand,
\be \label{SeffSYK}
S_{eff}/N= \frac{1}{4} g\star\( K_c^{-1} - (q-1)J^2\) \star g~.
\ee
The four-point function $\langle \tG(\tau_1, \tau_2) \tG(\tau_3, \tau_4)\rangle$ computed with (\ref{SeffSYK}), after doing the Gaussian integral, reproduces  Eq.~\ref{Fab} for one flavor. 
\newline

Now consider the generalized model, (\ref{SeffApp}), for two flavors with $q_1=q_2=q$ and $\kappa_1 = \kappa_2 = 1/2$. From (\ref{SeffApp}), 
\be \label{Seff22}
\frac{2 S_{eff}}{N} = - \frac{1}{2} \log \det(\partial_{\tau} - \tS_1) - \frac{1}{2} \log \det(\partial_{\tau} - \tS_2) 
+ \frac{1}{2}\int d\tau_1 d\tau_2\( \tS_1 \tG_1+ \tS_2 \tG_2 - \frac{2 J^2}{q^2} \tG_1^q \tG_2^q\)
\ee
By symmetry, the saddle point is,
\be
G_1 = G_2\equiv G~, \ \ \ \ \ \Sigma_1 = \Sigma_2\equiv \Sigma = \frac{2 J^2}{q} G(\tau_1, \tau_2)^{2q-1}~.
\ee
As noted in Sec.~\ref{EA}, the saddle point equation is the same as the equation for SYK with a $2 q$ body interaction. Let us now study fluctuations about the saddle, $\tG_a = G + |G|^{1-q} g_a$, and
$\tS_a = \Sigma + |G|^{q-1} \sigma_a$. Expanding (\ref{Seff22}) to second order, 
\begin{multline}
\frac{2 S_{eff}}{N} = \frac{1}{2} \int d\tau_1 d\tau_2\, \[g_1(\tau_1, \tau_2) \sigma_1(\tau_1, \tau_2) + g_2(\tau_1, \tau_2) \sigma_2(\tau_1, \tau_2) \right. \\
\left.- \frac{J^2}{ q} (q-1)( g_1(\tau_1, \tau_2)^2+ g_2(\tau_1, \tau_2)^2) - 2 J^2 g_1(\tau_1, \tau_2) g_2(\tau_1, \tau_2)\]\\  -\frac{1}{4} \int d\tau_1 \ldots d\tau_4\, K_{\bar{c}}(\tau_1, \ldots, \tau_4) \Big(\sigma_1(\tau_1, \tau_2) \sigma_1(\tau_3, \tau_4) + \sigma_2(\tau_1, \tau_2) \sigma_2(\tau_3, \tau_4)\Big)~,
\end{multline}
where 
\be
K_{\bar{c}}(\tau_1, \ldots, \tau_4) =   -|G(\tau_1, \tau_2)|^{q-1}G(\tau_1, \tau_3) G(\tau_2, \tau_4) |G(\tau_3, \tau_4)|^{q-1}~.
\ee
Integrating out $\sigma_1,\sigma_2$ gives, 
 \be
 \frac{ S_{eff}}{N} = \frac{1}{4} g_S \star\(K_{\bar{c}}^{-1} - \frac{2 J^2}{q}(2q-1)\) \star g_S + \frac{1}{4} g_A \star\(K_{\bar{c}}^{-1} - \frac{2 J^2}{q}(-1)\)\star g_A~,
 \ee
 where $g_S/g_A$ are the symmetric/antisymmetric combinations of $g_1, g_2$, (\ref{gsa}). The correlators $\langle \tG_S(\tau_1, \tau_2) \tG_S(\tau_3, \tau_4)\rangle$ and $\langle \tG_A(\tau_1, \tau_2) \tG_A(\tau_3, \tau_4)\rangle$  follow by analogy with (\ref{SeffSYK}), and reproduce (\ref{gsgs}) and (\ref{gaga}). 
 
 \section{Model with a Scalar} \label{AppendixScalar}
In this appendix we consider a model with a boson field. It is a slight variant of (\ref{GFM}) and has the action, 
\be \label{GFM2}
S= \int d \tau\,  \( \sum_{i=1}^{N_1} \phi_i^2+  \frac{1}{2}\sum_{a =2}^f \sum_{i=1}^{N_a}\chi_i^a \frac{d }{d \tau} \chi_i^a +\frac{(i)^{\frac{\mathrm{q}}{2}}}{\prod_{a=1}^f q_a!} \sum_I J_{I}\, \phi_i ( \chi_{i_1}^2 \cdots \chi_{i_{q_2}}^2)\cdots (\chi_{j_1}^f \cdots \chi_{j_{q_f}}^f) \)~,
\ee
where $I$ is a collective site index $I =i,  i_1,\ldots, i_{q_2}, \ldots, j_{1},\ldots, j_{q_f}$, and $q_1=1$, and $\mathrm{q} = \sum_{a=2}^f q_a$. This has a similar interaction as (\ref{GFM}), but the first flavor is with a boson instead of a fermion. The boson is taken to be auxiliary, having UV dimension $\[\phi\]=1/2$, so the coupling $J_I$ has dimension $1/2$. We restrict to only one boson, $q_1 =1$,  in order to ensure that the interaction is relevant. We also require  $\mathrm{q}$ be even.~\footnote{
A supersymmetric variant of SYK was introduced in \cite{Anninos:2016szt} (see also \cite{Anninos:2013nra}). A  more minimal supersymmetric SYK,  with only the interaction $\sum J_{i j k} \phi_i \chi_j \chi_k$, is being studied in \cite{Yu, FGMS} .  This interaction would be a special case of (\ref{GFM2}) with one fermion flavor with $q_2=2$. }

The only technical distinction between finding the IR dimensions for (\ref{GFM2}) compared with (\ref{GFM}) is that the boson propagator is symmetric in time.
The ansatz for the IR boson propagator is, 
\be
G_1 (\tau) = b_1 \frac{J^2}{|J^2 \tau|^{2\Delta_1}}~, \ \ \ \ \ G_1(\omega) = - 2 i \Delta_1 b_1 J^{2-4\Delta_1} |\omega|^{2 \Delta_1 -1} \psi(\Delta_1 + \frac{1}{2})~,
\ee
where $\psi(\Delta)$ is defined in (\ref{psiDelta}), while  for the IR fermion propagator it is, 
\be
G_k (\tau) = b_k \frac{\sgn(\tau)}{|J^2 \tau|^{2\Delta_k}}~, 
\ee
for $k \geq 2$.  In the IR, one drops the free propagator appearing in the Schwinger-Dyson equation (\ref{SD}), so for both the boson and the fermions one has $\Sigma_k(\omega) G_k(\omega) = -1$. The self-energy is again given by (\ref{SigGen}), with the disorder average normalization given in (\ref{DisAvgGF}). Repeating the several steps in Sec.~\ref{Sec:Gen} gives the following equations, 
\bea \label{bsyk1}
\sum_{a=1}^f q_a \Delta_a = 1, \ \ \ \ \prod_{a=1}^f b_a^{q_a} &=& \frac{  \kappa_k Q_k}{2\pi} (1-2\Delta_k ) \tan \pi \Delta_k  \ \ \text{for } k\in\{2,\ldots, f\}\\ \label{bsyk2}
\prod_{a=1}^f b_a^{q_a} &=& \frac{  \kappa_1 Q_1}{2\pi} \frac{(1- 2\Delta_1 )}{\tan \pi \Delta_1}
\eea
The equations (\ref{bsyk1}) are the same as for the generalized fermion model (\ref{GFM}), while (\ref{bsyk2}) is different. 

For the case of $f=2$ and $\kappa_1 = \kappa_2$, (\ref{bsyk1}, \ref{bsyk2})  have the simple solution, 
\be
\Delta_1 = \frac{q_2+2}{2 q_2+2}~, \ \ \ \ \ \ \Delta_2 =\frac{1}{2 q_2+2}~.   
\ee
For $q_2=2$ this gives the dimensions $\Delta_1 = 2/3$, $\Delta_2 =1/6$ found in \cite{Yu, FGMS}.~\footnote{We thank Yu Nakayama for sharing his results and explaining the supersymmetric model.} Intriguingly, the difference between the boson dimension $\Delta_1$ and fermion dimension $\Delta_2$ is $1/2$ for any $q_2$, as would be implied by supersymmetry. However, we have not checked that the model for general $q_2$ is supersymmetric.

\section{Random Mass Matrix Fermions} \label{AppendixA}
The simplest SYK model is for $q=2$:  fermions with a random mass matrix. For this case, all computations can be performed exactly, without the restriction of being near the fixed points or working at large $N$. In this appendix, we solve the $q=2$ model. For  infinite $N$ this is trivial, while for finite $N$ it is slightly more involved but follows from standard matrix model techniques. One should keep in mind that the $q=2$ case has multiple features that are not representative of SYK at larger $q$; in particular, it is not chaotic.~\footnote{In \cite{Magan} it was proposed that the $q=2$ model satisfies the Eigenstate Thermalization Hypothesis.}

\subsection{Infinite N}
\begin{figure}[t]
\centering
\includegraphics[width=4in]{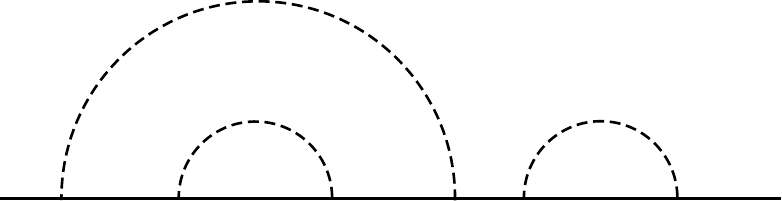}
\caption{The  model of fermions with a random mass matrix sums rainbow diagrams like the one shown. The solid line is the fermion and the dashed line is the disorder.} \label{Rainbow}
\end{figure}
The Schwinger-Dyson  equations for the two-point function (\ref{SD}, \ref{SD2})  are integral equations for general $q$, but become a simple quadratic equation for $q=2$. The solution is
\be \label{Gq2v1}
G(\omega) = \frac{i \omega}{ 2 J^2}\( -1 + \sqrt{1+ 4\frac{J^2}{\omega^2}}\)~.
\ee
One can also find this directly by summing non-crossing rainbow diagrams (see Fig.~\ref{Rainbow}),
 \be \label{G342}
G(\omega) = G_0(\omega) \sum_{n=0}^{\infty} C_{n} \Big(J G_0(\omega)\Big)^{2n}~,
\ee
where $G_0(\omega)$ is the bare propagator (\ref{Gweak}), while $C_n$ are the Catalan numbers, 
\be
C_n = \frac{1}{n+1} \binom{2n}{ n}~.
\ee
The Catalan numbers are the number of different ways $n+1$ factors can be completely parenthesized; here the parentheses are the rainbows. Summing (\ref{G342}) gives (\ref{Gq2v1}). One can also write (\ref{Gq2v1}) as, 
\be \label{Gspec}
G(\omega) = \int d\lambda\, \rho(\lambda)\, \frac{-1}{i \omega - \lambda}~, 
\ee
where the spectral function $\rho(\lambda)$ is the Wigner semi-circle, with support  for $|\lambda|< 2J$,
\be \label{Wigner}
\rho(\lambda) = \frac{1}{2\pi J^2} \sqrt{4J^2 - \lambda^2}~.
\ee

At finite temperature, the frequencies in (\ref{Gq2v1}) should be viewed as the Matsubara frequencies, $\omega_{n} = (2n+1) \pi/\beta$. Taking the discrete Fourier transform of (\ref{Gspec}) gives, 
\be
G_{\beta}(\tau) = \int d\lambda\, \rho(\lambda)\, \frac{1}{1+e^{-\beta \lambda}}\, e^{-\lambda \tau}~,
\ee
where $0< \tau<\beta$. 
In the limit of zero temperature $\beta J \gg 1$, we can evaluate the integral to obtain, 
\be \label{Gq2t}
G(\tau) = \frac{1}{2 J\tau}\Big( I_1 (2 J |\tau|)-  \sL_1 (2 J |\tau|) \Big)~,
\ee
where $\sL_1$ is the modified Struve function, and $I_1$ is the modified Bessel function. While both $\sL_1$ and $I_1$  grow exponentially, the two-point function (\ref{Gq2t}) decays monotonically with $|J\tau|$. The combination is sometimes denoted by $\sM_1 \equiv \sL_1 - I_1$. 
We can do a strong coupling expansion of (\ref{Gq2t}), 
\be \label{Gq2strong}
G(\tau) = -\frac{1}{2\pi^2 J \tau} \sum_{k=0}^{\infty} \frac{\Gamma\(k+\frac{1}{2}\) \Gamma\(k-\frac{1}{2}\)}{|J \tau|^{2 k}}~,
\ee
where we see that the first term matches what was expected from the IR limit of the Schwinger-Dyson equations (\ref{Gstrong}, \ref{DeltaSYK}). 

\subsection*{Comments}
One comment is that in summing the Feynman diagrams giving (\ref{G342}) it is important to work at finite temperature. Each of the  diagrams individually has IR divergences:  the Fourier transform of any of the individual terms in (\ref{G342}) will diverge in the limit of $\beta\rightarrow \infty$. Of course, one could have chosen to regulate the IR divergence in some way other than working at finite temperature. However, finite temperature is natural. The point is just that the dimensionless expansion parameter is $\beta J$.

Another comment is that aside from the implicit appearance of $\beta$ in the Matsubara frequencies, (\ref{G342}) has no explicit $\beta$ dependance. This is a property that is special to $q=2$. For SYK with $q \geq 4$, one could solve the Schwinger-Dyson equations perturbatively around weak coupling, giving an expansion of the form, 
\be
G(\omega_n) = \frac{i}{\omega_n} \sum_{k=0}^{\infty} \sum_{l=0}^k g_{k l} \(\frac{J}{\omega_n}\)^{2 k} \(\omega_n \beta\)^{2 l}~,
\ee
with some coefficients $g_{ k l}$. One can derive recursion relations for $g_{k l}$, but we have not found a way of solving them.

We have been considering Majorana fermions. One can instead study $q=2$ with Dirac fermions, 
\be \label{HD2}
H = \sum_{i j} J_{i j} c_i^{\dagger} c_j~.
\ee
At leading order in $1/N$, this gives the same two-point function (\ref{Gq2v1}). Working with Dirac fermions gives slightly more flexibility, as one can introduce a chemical potential. With no chemical potential, as in (\ref{HD2}), one is at half-filling. 
Explicitly, consider a single free Dirac fermion $H = \omega_0 c^{\dagger} c$. (Adding a chemical potential just corresponds to adding to (\ref{HD2}) such a term for each fermion, with chemical potential $\mu = - \omega_0$.) The finite-temperature two-point function is trivially, 
\be \label{GDirac}
G(\tau) = Z^{-1} \tr( T e^{- \beta H}\, c(\tau) c^{\dagger}(0)) = \frac{1}{1 + e^{- \beta \omega_0}} \( \theta(\tau) e^{- \omega_0 \tau} - \theta(-\tau) e^{- \beta \omega_0 - \omega_0 \tau}\)~.
\ee
Since we are at finite temperature, fields have the time range $0<\tau<\beta$. The two-point function is a function of the difference between two times, and so naturally has the range $-\beta<\tau<\beta$.  However, from (\ref{GDirac}) we see that for $0<\tau<\beta$, $G(\tau - \beta) = - G(\tau)$. We can thus restrict to $0<\tau<\beta$. The filling fraction $\mathcal{Q}$ is defined as the expectation value of the occupation number,
\be
\mathcal{Q} = \langle c^{\dagger} c\rangle = -G(0^{-}) = \frac{1}{1+e^{\beta \omega_0}}~.
\ee
We can choose the filling fraction by choosing $\omega_0$. It is clear that for any finite temperature, if $\omega_0=0$, then there is no energy cost to being in the state $|1\rangle$ versus $|0\rangle$, and so the filling fraction is $1/2$. Note that the limits of $T\rightarrow 0$ and $\omega_0\rightarrow 0$ do not commute. If we set $T=0$ at finite $\omega_0$ (including $\omega_0=0$), then we get zero filling: from (\ref{GDirac}), $G(\tau) = \theta(\tau)$. 

Finally, the $q=2$ model sums rainbow diagrams. There are many other models that sum rainbow diagrams. For instance,  two-dimensional QCD \cite{t1974two} has the same rainbow diagrams, where  the fermions are the quarks, and the disorder lines are the gauge field propagators. Also, the recently studied three-dimensional $U(N)_k$ Chern-Simons theory coupled to scalars or fermions also sums rainbow-like diagrams \cite{Aharony:2011jz, Giombi:2011kc}. A simple large $N$ quantum mechanics model that sums rainbow diagrams is the Iizuka-Polchinski model \cite{IP} (see also, \cite{IOP, MPRS}). The IP model has a harmonic oscillator in the adjoint representation of $U(N)$ plus a harmonic oscillator in the fundamental representation of $U(N)$, coupled through a trilinear interaction. In the limit that the mass of the adjoint goes to zero, this is essentially the same as the model Eq.~\ref{HD2}, at leading order in $1/N$. The reason we say essentially the same is because in the IP model the fundamental is effectively at zero filling. In other words, its free two-point function is $\theta (\tau)$ as opposed to $\frac{1}{2}\sgn(\tau)$, and correspondingly, the infinite $N$ two-point function after summing the rainbow diagrams is only the first term, $I_1$, in (\ref{Gq2t}). In addition, at subleading orders in $1/N$, differences will arise between  the model Eq.~\ref{HD2} and the IP model. This is because the adjoint propagator will receive quantum corrections, whereas the $\langle J_{i j} J_{i j}\rangle$ ``propagator'' in  Eq.~\ref{HD2} is always a constant.

\subsection*{Four-Point Function}
It is simplest to write the four-point function in frequency space. This is defined as,
\be
\mF_{i j k l}(\omega_1, \omega_2, \omega_3,\omega_4) \equiv \int d\tau_1\ldots d\tau_4\, e^{i (\omega_1 \tau_1 + \ldots + \omega_4 \tau_4)} \,\langle \chi_i(\tau_1) \chi_j(\tau_2) \chi_k(\tau_3)\chi_l(\tau_4)\rangle~.
\ee
Written as a series in $1/N$, $\mF_{i j k l} = \mF_{i j k l }^{(0)} + \frac{1}{N} \mF_{i j k l }^{(1)}+ \ldots$~.  At leading order in $1/N$ there is a disconnected piece,
\be\nonumber
\!\!\!\!\mF_{i j k l}^{(0)} =\delta_{i j} \delta_{k l} 2\pi \delta(\omega_1 + \omega_2) 2\pi \delta(\omega_3 + \omega_4)\, G(\omega_1) G(\omega_3)
 - \( j \leftrightarrow k, \omega_2 \leftrightarrow \omega_3\) -\(j \leftrightarrow l, \omega_2 \leftrightarrow \omega_4\).
\ee
At first subleading order in $1/N$, the four-point function is a sum of ladder diagrams, like SYK for general $q$. However for $q=2$ there is an extreme simplification, since the rungs only contain the disorder lines. Since there is no momentum exchange, summing the ladders in frequency space simply involves summing a geometric series,  which gives, 
\begin{multline} \label{4ptNInf}
\!\!\!\!\!\! \mF_{i j k l}^{(1)} = \(\delta_{i l} \delta_{j k} \frac{J^2 \(G(\omega_2) G(\omega_4)\)^2}{1 - J^2 G(\omega_2) G(\omega_4)} - ( k \leftrightarrow l, \omega_3 \leftrightarrow \omega_4)\) 2\pi \delta(\omega_1 + \omega_2) 2\pi\delta(\omega_3 + \omega_4)\\
- \Big( j \leftrightarrow k, \omega_2 \leftrightarrow \omega_3\Big) - \Big(j \leftrightarrow l, \omega_2 \leftrightarrow \omega_4\Big)~.
\end{multline}
It is only necessary to establish the first term in the $s$-channel piece. The other term, as well as the $t$ and $u$ channels, follow from antisymmetry. 
Through a Fourier transform and analytic continuation of (\ref{4ptNInf}), one finds there is no exponential growth in the out-of-time-order four-point function \cite{MPRS}, and so the $q=2$ model is not chaotic.

\subsection{Finite N} \label{FiniteN}
\subsubsection{Dirac Fermion}
We now compute the two-point function at finite $N$ for the random mass matrix fermion, (\ref{HD2}). For fixed coupling $J_{i j}$, this is just $N$ free fermions with mass matrix $J_{i j}$, so the nontrivial part is to perform the disorder average. Specifically, 
\be \label{2ptN}
G(\omega ) = -\frac{1}{N} \frac{1}{Z} \int \prod_{i\leq j}d J_{i j}\, \text{tr}\( \frac{1}{i\omega - J}\)\, \exp\(-\text{tr}(J^2)/2 \bar{J}^2\)~,
\ee
where,~\footnote{We are using the same symbol $J$ to denote both the matrix of couplings, as well as the number that appears as the variance of the distribution of couplings.}
\be \label{Norm}
Z = \int \prod_{i\leq j} d J_{i j}\, \, \exp\(-\text{tr}(J^2)/ 2 \bar{J}^2\)~, \ \ \ \ \ \ \ \bar{J}^2 = \frac{J^2}{N}~.
\ee
Consider first the trivial case of $N=1$. This is just a fermion with a random mass. Then (\ref{2ptN}) reduces to  (\ref{Gspec}) with a spectral function, 
\be \label{N1spec}
\rho(\lambda) = \frac{1}{\sqrt{2\pi} J} e^{- \frac{\lambda^2}{2 J^2}}~.
\ee
So at $N=1$ the spectral function is a Gaussian, while at $N=\infty$ it is the Wigner semicircle (\ref{Wigner}). The two-point function for $N=1$ can also be found by summing Feynman diagrams, 
\be \label{GG22}
G(\omega) = G_0(\omega) \sum_{n=0}^{\infty} (2n-1)!! \(J G_0(\omega)\)^{2n}~.
\ee
 Writing the double factorial as a Gaussian integral, and interchanging the integral and the sum, we recover (\ref{N1spec}). Explicitly performing the integral gives the two-point function in terms of the complimentary error function, 
\be \label{GoldO}
G(\omega) =\frac{i}{J} \sqrt{\frac{\pi}{2}}\, e^{\frac{\omega^2}{2J^2}}\, \text{Erfc} \(\frac{\omega}{\sqrt{2} J}\)~, \ \ \ \ \omega>0~.
\ee
In the zero temperature limit we also find, 
\be \label{27}
G(\tau) = \frac{1}{2} e^{\frac{J^2 \tau^2}{2}} \text{Erfc}\( \frac{ J \tau}{\sqrt{2}}\)~, \ \ \ \ \ \ \ \tau>0~,\ \beta J\gg 1~.
\ee
We now move on to the case of general $N$, using the method of orthogonal polynomials to evaluate (\ref{2ptN}). We can write (\ref{Norm}) in terms of the eigenvalues of $J$, 
\be \label{eZ}
Z = \int  \prod_{k=1}^{N} d\lambda_k\, \prod_{1\leq i< j\leq N}\!\!\! \( \lambda_i - \lambda_j\)^2\, e^{ - \lambda_l^2/ 2 \bar{J}^2}= \int  \prod_{k=1}^{N} d\lambda_k\,  \Delta(\lambda)^2 e^{ - \lambda_i^2/ 2 \bar{J}^2}~,
\ee
where we have used that $J$ is Hermitian and the last equation is in terms of the Vandermonde,
\be \label{van}
\Delta(\lambda) = \begin{vmatrix}
1 & \lambda_1 & \lambda_1^2 & \ldots & \lambda_{1}^{N-1} \\
1 & \lambda_2 & \lambda_2^2 & \ldots & \lambda_2^{N-1}\\
\vdots & \vdots  &\vdots & \ldots &\vdots\\
1 & \lambda_{N} & \lambda_{N}^2 & \ldots & \lambda_{N}^{N-1}
\end{vmatrix}~.
\ee
This model is of course different from fermions with masses independently drawn from a Gaussian distribution; the masses here are eigenvalues of a Hermitian matrix and have repulsion, as encoded in the Vandermonde term in (\ref{eZ}). 

\begin{figure}[t]
\centering
\includegraphics[width=3in]{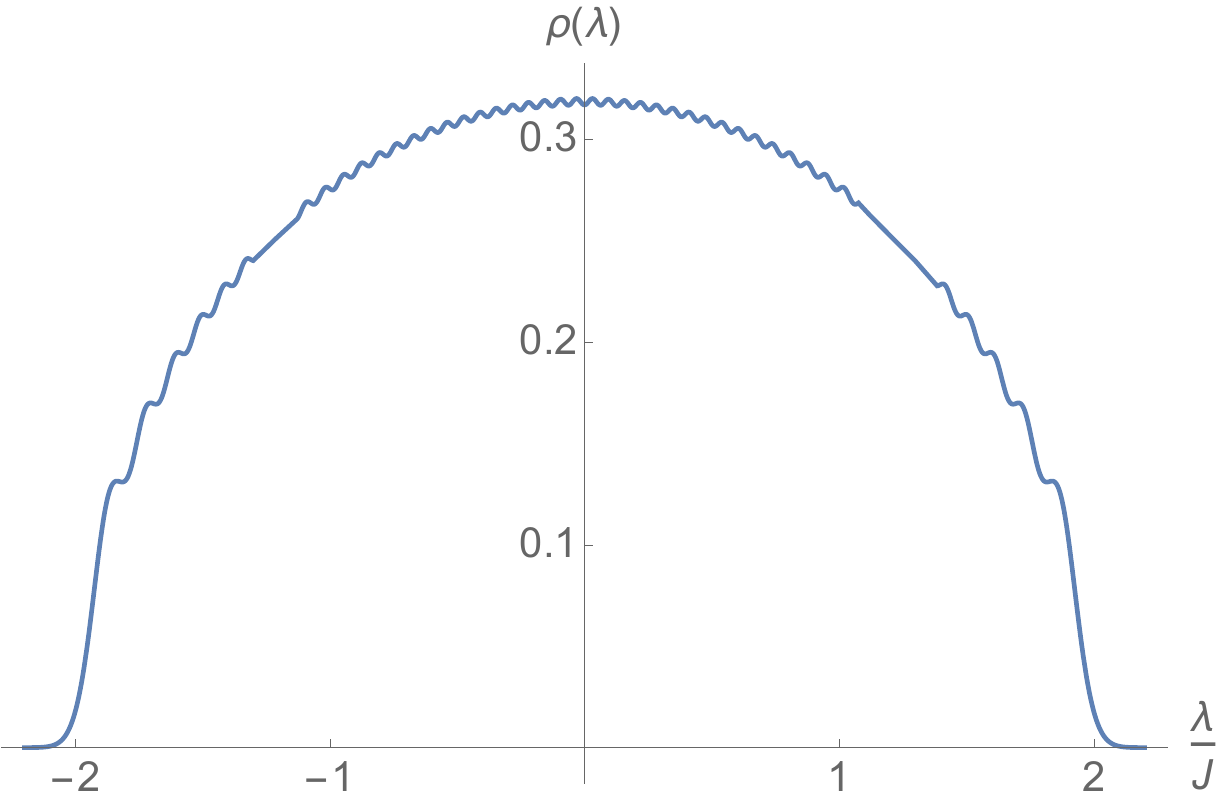}
\caption{Plot of the spectral function (\ref{rhoLambda}) at $N=50$ for the random mass matrix fermion (\ref{HD2}). At infinite $N$ the oscillations go away and this becomes the Wigner semi-circle (\ref{Wigner}).} \label{rhoPlot}
\end{figure}

We now take linear combinations of the columns of (\ref{van}), transforming it into a matrix with $i,j$ element, $\phi_{j}(\lambda_i)$, where $\phi_{j}(\lambda_i)$ is a polynomial with lowest element $1$ and highest element $\lambda_i^{ j}$. 
The determinant (\ref{van}) remains invariant under these operations. We can write the determinant as a sum of permutations of the  integers from $0$ to $N-1$, 
\be \label{van2}
\Delta(\lambda) = \sum_{\sigma} (-)^{\sigma} \phi_{\sigma(0)} (\lambda_1) \phi_{\sigma(1)}(\lambda_2) \cdots \phi_{\sigma(N-1)}(\lambda_{N})~.
\ee
We choose the $\phi_n$ such that, 
\be \label{van3}
\int d\lambda\, \phi_n(\lambda) \phi_m(\lambda)\, e^{- \lambda^2/ 2 \bar{J}^2} = f_n\, \delta_{n m}~.
\ee
This then gives, 
\be 
Z = N! \prod_{i=0}^{N-1} f_{i}~.
\ee
The $\phi_n$ will be proportional to the Hermite polynomials, defined as,
\be \label{orth}
H_n(x) = (-1)^n e^{x^2} \frac{d^n}{dx^n} e^{-x^2}~, \ \ \ \ \ \int_{-\infty}^{\infty} dx\ e^{-x^2} H_n(x) H_m(x) = \sqrt{\pi} 2^n n!\, \delta_{n m}~.
\ee
We choose,  
\be \label{phin}
\phi_n(\lambda) = \frac{\bar{J}^n}{2^{n/2}} H_n \(\frac{\lambda}{\sqrt{2} \bar{J}}\)~.
\ee
Now to evaluate (\ref{2ptN}), note that, 
\be
\text{tr}\(\frac{1}{i\omega - J}\) =  \sum_{i=1}^{N} \frac{1}{i\omega - \lambda_i}~,
\ee
and so we find that the spectral function is,
\be \label{specN}
\rho(\lambda) = \frac{1}{N}e^{- \frac{\lambda^2}{2 \bar{J}^2}} \sum_{k=0}^{N-1} \frac{\phi_k(\lambda)\phi_k(\lambda)}{f_k} ~.
\ee
Evaluating the sum gives, 
\be \label{rhoLambda}
\!\!\!\!\!\!\! \rho(\lambda) =\frac{1}{\sqrt{2\pi} \bar{J}}\frac{1}{2^N N!} \, e^{-\frac{\lambda^2}{2 \bar{J}^2}} \[ N H_{N-1}\(\frac{\lambda}{\sqrt{2} \bar{J}}\)^2 - (N-1) H_{N-2}\(\frac{\lambda}{\sqrt{2} \bar{J}}\)H_{N}\(\frac{\lambda}{\sqrt{2} \bar{J}}\)\]
\ee
where we have used that $f_{k} = \bar{J}^{2k+1} \sqrt{2\pi}\, k!$, which follows from (\ref{orth}, \ref{phin}). A plot of (\ref{rhoLambda}) is shown in Fig.~\ref{rhoPlot}.

An alternative way to write the two-point function is to perform the integral over $\lambda$ in (\ref{Gspec}) before evaluating the sum over $k$ appearing in the spectral function (\ref{specN}). After the introduction of a Schwinger parameter, the integration over $\lambda$ yields a Laguerre polynomial. Using that the sum of the Laguerre polynomials is an associated Laguerre polynomial $\sum_{k=0}^{N-1} L_k(x) = L_{N-1}^1(x)$, we find
\be \label{Gdl}
G(\omega) = \frac{i}{\omega}\frac{1}{N} \int_0^{\infty} ds\, e^{-s} e^{ - \frac{s^2 \bar{J}^2}{2 {\omega}^2}}\, L_{N-1}^{1} \(\frac{s^2 \bar{J}^2}{\omega^2}\)~.
\ee
\subsection*{1/N expansion}
We would like to expand (\ref{Gdl}) in powers of $1/N$. Using the definition of the associated Laguerre polynomial,
\be
L_n^{\alpha}(x) = \sum_{k=0}^{n} (-1)^k \binom{n+\alpha}{n-k}~ \frac{x^k}{k!}~,
\ee
and recalling that $\bar{J}^2 \equiv J^2/N$, we exchange the order of the sums, and perform the integral over $s$, to get,\be
G(\omega) = \frac{i}{\omega} \sum_{p=0}^{\infty} (-1)^p \(\frac{J}{\omega}\)^{2p} \frac{ (2p)!}{p!\, (p+1)!}  B(p,N)~,
\ee
where \cite{DG}
\be
B(p,N) = \frac{(p+1)!}{(2N)^p} \sum_{k=0}^{p} \frac{2^k (N-1)!\, p!}{(N-1-k)!\, k! \,(k+1)!\, (p-k)!} = \frac{(p+1)!}{(2N)^p}\, \phantom{}_2 F_1(-p, 1-N;2;2)~.
\ee
An series expansion of  $B(p,N)$ in  powers of $1/N^2$ was also worked out in \cite{DG}. The first few terms are,
\be
B(p,N) = 1 + \frac{p(p^2 -1)}{12 N^2} + \frac{(p+1)! }{(p-4)!}\frac{(5p-2)}{1440 N^4} + \frac{(p+1)!}{(p-6)!} \frac{(35 p^2 - 77 p + 12)}{2^7 3^4 5 \cdot 7 N^6} + \ldots~,
\ee
Using this we can write the $1/N$ expansion of the two-point function as, 
\be \label{Nexp}
G(\omega) = \frac{i}{\omega}\sum_{n=0}^{\infty} N^{-2n}\, g^{(n)}\!\!\(\frac{J}{\omega}\),
\ee
where the first several terms are,
\bea \nonumber
g^{(0)}(x) &=& \frac{-1 + \sqrt{1 + 4 x^2}}{2 x^2}~,\\ \nonumber
g^{(1)}(x) &=& \frac{x^4}{(1 + 4 x^2)^{\frac{5}{2}}}~,\\ \nonumber
g^{(2)}(x) &=& -\frac{21 x^8 (x^2-1)}{(1 + 4 x^2)^{\frac{11}{2}}}~,\\ \nonumber 
g^{(3)}(x) &=& \frac{11\, x^{12} (158 x^4 - 558 x^2 +135)}{(1 + 4 x^2)^{\frac{17}{2}}}~.
\eea
The leading term in $1/N$, $g^{(0)}$, reproduces what we found from summing the planar diagrams, (\ref{Gq2v1}).

\subsubsection{Majorana Fermion}
Here we compute the two-point function for the Majorana version of $q=2$ SYK (\ref{SYK}) at finite $N$ (note that $N$ must be even).
This will be slightly different from the Dirac version studied in Sec.~\ref{FiniteN}. 
The two-point function is given by,
\be \label{2ptNM}
G(\omega ) = -\frac{1}{N}\frac{1}{Z} \int \prod_{i<j}d J_{i j}\, \text{tr}\( \frac{1}{i\omega - J}\)\, \exp\(-\text{tr}(J^2)/4 \tilde{J}^2\)~,
\ee
where
\be \label{NormM}
Z = \int \prod_{i<j} d J_{i j}\, \, \exp\(-\text{tr}(J^2)/ 4 \tilde{J}^2\)~, \ \ \ \ \tilde{J}^2 = \frac{J^2}{N-1}~.
\ee
The matrix $J$ is real and antisymmetric. The partition function (\ref{NormM}) can be written terms of the eigenvalues of $J$ \cite{Mehta},
\be \label{eZM}
Z = \int  \prod_{k=1}^{N/2} d\lambda_k\, \prod_{1\leq i< j\leq N/2}\!\!\! \( \lambda_i^2 - \lambda_j^2\)^2\, e^{ - \lambda_l^2/ 2 \tilde{J}^2}~.
\ee
Defining an analog of the Vandermonde, one involving only even powers,
\be \label{vanM}
\Delta(\lambda) = \begin{vmatrix}
1 & \lambda_1^2 & \lambda_1^4 & \ldots & \lambda_{1}^{N-2} \\
1 & \lambda_2^2 & \lambda_2^4 & \ldots & \lambda_2^{N-2}\\
\vdots & \vdots  &\vdots & \ldots &\vdots\\
1 & \lambda_{N/2}^2 & \lambda_{N/2}^4 & \ldots & \lambda_{N/2}^{N-2}
\end{vmatrix}~,
\ee
Eq.~\ref{eZM} becomes, 
\be
Z = \int  \prod_{k=1}^{N/2} d\lambda_k\,  \Delta(\lambda)^2 e^{ - \lambda_i^2/ 2 \tilde{J}^2}~.
\ee
The procedure is now similar to the Dirac case. We can write the determinant as a sum of permutations of the even integers from $0$ to $N-2$, 
\be \label{van2}
\Delta(\lambda) = \sum_{\sigma} (-)^{\sigma} \phi_{\sigma(0)} (\lambda_1) \phi_{\sigma(2)}(\lambda_2) \cdots \phi_{\sigma(N-2)}(\lambda_{N/2})~.
\ee
The $\phi_n$ are the same as in the Dirac case. The partition function now involves just the even normalization constants, 
\be 
Z = (\frac{1}{2}N)! \prod_{i=0}^{\frac{N}{2}-1} f_{2i}~.
\ee
For evaluating the two-point function, note that since the eigenvalues come in pairs, 
\be
\text{tr}\(\frac{1}{i\omega - J}\) =  \sum_{i=1}^{N/2}\( \frac{1}{i\omega + \lambda_i} + \frac{1}{i\omega - \lambda_i}\)~.
\ee
The two-point function is thus,
\be \label{GMfiniteN}
G(\omega) = \frac{ i}{\omega} \frac{2}{N} \, \int_0^{\infty} d s\, e^{-s}\, e^{-\frac{ s^2 \tilde{J}^2}{2 \omega^2}}\,  \sum_{k=0}^{ N-1} \frac{1+(-1)^k}{2} L_{k}\(\frac{s^2 \tilde{J}^2}{\omega^2}\)~.
\ee
This is similar to (\ref{Gdl}), except it involves a sum only over the even Laguerre's. 
\begin{figure}[t]
\centering
\includegraphics[width=3.2in]{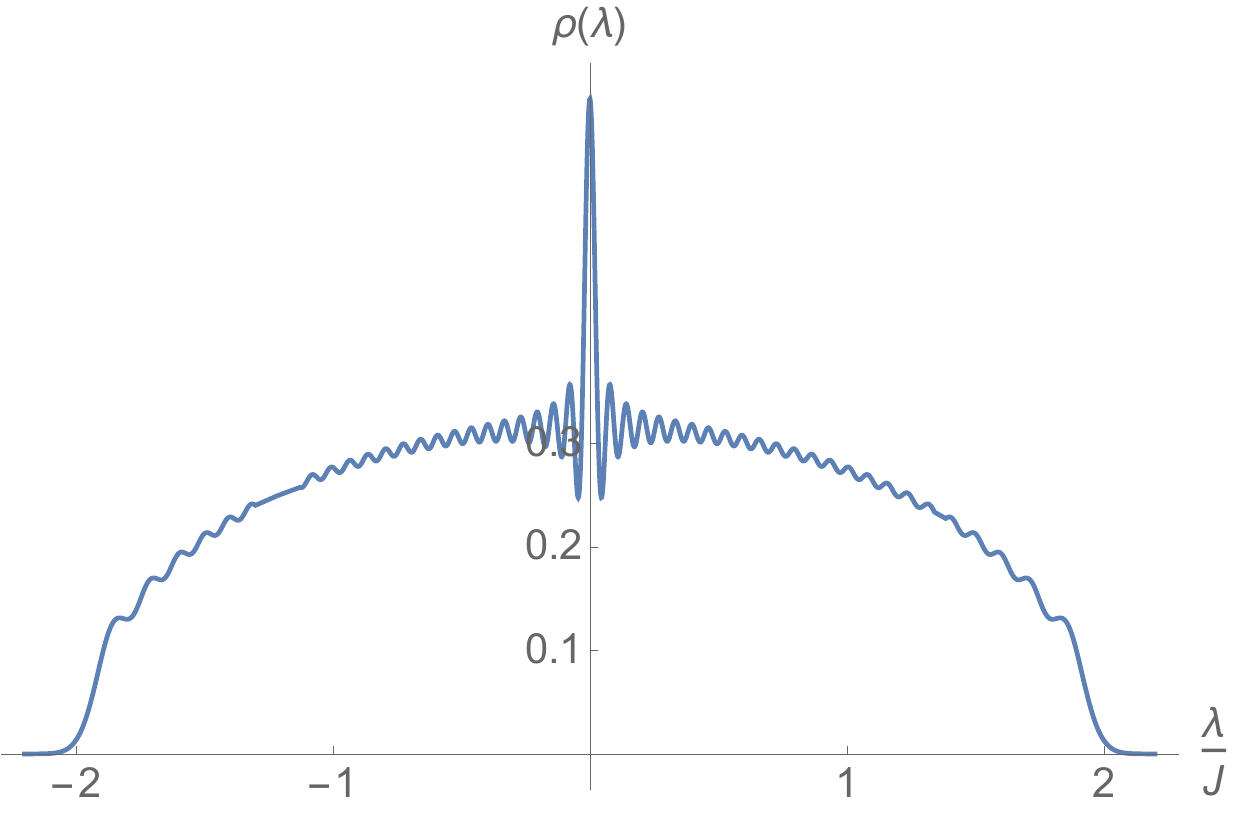}
\caption{Plot of the spectral function (\ref{rhoLambdaM}) at $N=50$ for the $q=2$ Majorana SYK. This differs from the random mass matrix fermion spectral function in the region of small $\lambda$, see Fig.~\ref{rhoPlot}; the distinction goes away at infinite $N$. } \label{MajoranaPlot}
\end{figure}
The spectral function is, 
\begin{multline} \label{rhoLambdaM}
\rho(\lambda) =\frac{1}{\sqrt{2\pi} \tilde{J}}\frac{1}{2^N N!} \, e^{-\frac{\lambda^2}{2 \tilde{J}^2}} \[ N H_{N-1}\(\frac{\lambda}{\sqrt{2} \tilde{J}}\)^2  - (N-1) H_{N-2}\(\frac{\lambda}{\sqrt{2} \tilde{J}}\)H_{N}\(\frac{\lambda}{\sqrt{2} \tilde{J}}\) \right.\\
\left. - \frac{\tilde{J}}{\sqrt{2} \lambda} H_{N}\(\frac{\lambda}{\sqrt{2} \tilde{J}}\) H_{N-1}\(\frac{\lambda}{\sqrt{2} \tilde{J}}\) \]~.
\end{multline}
This is similar to the spectral function for the Dirac fermion (\ref{rhoLambda}), except for the addition of the last term in (\ref{rhoLambdaM}) that is $1/N$ suppressed relative to the first two (and the trivial distinction that occurs at order $1/N$ between $\tilde{J}$ and $\bar{J}$).


\bibliographystyle{utphys}

\end{document}